\documentclass[twocolumn,english,aps,pra,showpacs,superscriptaddress]{revtex4-1}
\usepackage[T1]{fontenc}
\usepackage[latin9]{inputenc}
\usepackage{amsmath}
\usepackage{amssymb}
\usepackage{graphicx}

\makeatletter

\@ifundefined{textcolor}{}
{%
 \definecolor{BLACK}{gray}{0}
 \definecolor{WHITE}{gray}{1}
 \definecolor{RED}{rgb}{1,0,0}
 \definecolor{GREEN}{rgb}{0,1,0}
 \definecolor{BLUE}{rgb}{0,0,1}
 \definecolor{CYAN}{cmyk}{1,0,0,0}
 \definecolor{MAGENTA}{cmyk}{0,1,0,0}
 \definecolor{YELLOW}{cmyk}{0,0,1,0}
}

\usepackage{amsfonts}
\@ifundefined{definecolor}
 {\usepackage{color}}{}

\def\be{\begin{equation}}
\def\ee{\end{equation}}
\def\bea{\begin{eqnarray}}
\def\eea{\end{eqnarray}}
\def\bi{\begin{itemize}}
\def\ei{\end{itemize}}

\makeatother

\usepackage{babel}
\begin{document}

\title{Locally tunable disorder and entanglement  \\
in the one-dimensional plaquette orbital model }

\author{Wojciech Brzezicki}

\affiliation{Marian Smoluchowski Institute of Physics, Jagellonian University,
Reymonta 4, PL-30059 Krak\'ow, Poland}

\author{Andrzej M. Ole\'{s} }

\affiliation{Marian Smoluchowski Institute of Physics, Jagellonian University,
Reymonta 4, PL-30059 Krak\'ow, Poland}
\affiliation{Max-Planck-Institut f\"ur Festk\"orperforschung, Heisenbergstrasse 1,
D-70569 Stuttgart, Germany}

\date{3 May, 2014}

\begin{abstract}
We introduce a one-dimensional plaquette orbital model with a topology
of a ladder and alternating interactions between $x$ and $z$ pseudospin
components along both the ladder legs and on the rungs. We show that it
is equivalent to an effective spin model in a magnetic field, with spin
dimers that replace plaquettes and are coupled along the chain by
three-spin interactions. Using perturbative treatment and mean field
approaches with dimer correlations we study the ground state spin
configuration and its defects in the lowest excited states. By the
exact diagonalization approach we find that the quantum effects in the
model are purely short-range and we get estimated values of the ground
state energy and the gap in the thermodynamic limit from the system
sizes up to $L=12$ dimers. Finally, we study a class of excited states
with classical-like defects accumulated in the central region of the
chain to find that in this region the quantum entanglement measured by
the mutual information of neighboring dimers is locally increased and
coincides with disorder and frustration.
Such islands of entanglement in otherwise rather classical system
may be of interest in the context of quantum computing devices.
\end{abstract}


\maketitle

\section{Introduction}

Transition-metal oxides with active orbital degrees of freedom are
frequently described in terms of spin-orbital models
\cite{Tokura,Hfm,Ole05,Kha05,Ole12} which are realizations of the early
idea of Kugel and Khomskii \cite{Kug82} that orbital operators have to
be treated with their full dynamics in the limit of large on-site
Coulomb interactions. The interplay between spin and orbital
(pseudospin) interactions on superexchange bonds follows from the
mechanism of effective magnetic interactions at strong correlation and
is responsible for numerous quantum properties which originate from
spin-orbital entanglement \cite{Ole12}. This phenomenon is similar to
entanglement in spin models \cite{Ami08}, but occurs here in a larger
Hilbert space \cite{You12} and has measurable consequences at finite
temperature as found, for instance, in the phase diagrams
\cite{Hor08} and in ferromagnetic dimerized interactions \cite{Her11}
in the vanadium perovskites. In higher dimensional systems exotic spin
states are also triggered in the ground state by entangled spin-orbital
interactions in certain situations, as in:
(i) the $d^1$ spin-orbital model on the triangular lattice \cite{Cha11},
(ii) the two-dimensional (2D) Kugel-Khomskii model \cite{Brz12}, and
(iii) spinel and pyrochlore crystals with active $t_{2g}$ orbitals
\cite{Bat11}.
Such entangled spin-orbital states are very challenging but also
notoriously difficult to investigate except for a few exactly solvable
1D models \cite{Li98,Brz14}.

To avoid the difficulties caused by entanglement one considers
frequently ferromagnetic systems, where orbital interactions alone
are responsible for the nature of both the ground and excited states.
Orbital interactions in Mott insulators depend on the type of active
and partly filled $3d$ orbitals --- they have distinct properties for
either $e_g$ symmetry \cite{vdB99,vdB04,Fei05,Ryn10}, or $t_{2g}$
symmetry \cite{Dag08,Wro10,Tro13,Che13}. In contrast to spin models,
their symmetry is lower than SU(2) due to directional character of
orbital interactions which manifests itself in their intrinsic
frustration. The models which focus on such frustrated interactions
are the 2D compass model on the square lattice
\cite{vdB13,Nus05,Dou05,Dor05,Tan07,Wen08,Orus09,Dus09,Cin10,Brz10,Tro10},
the exactly solvable 1D compass model \cite{Brz07,You14}, the compass
ladder \cite{Brz09} and the Kitaev model on the honeycomb lattice
\cite{Kit06,Bas07}. The former includes only two spin components and
1D order arises in the highly degenerate ground state
\cite{Dou05,Dor05,Tan07,Wen08} which is robust with respect to
perturbing Heisenberg interactions \cite{Tro10}, while the latter
provides an exactly solvable case of a spin liquid with only nearest
neighbor (NN) spin correlations.

The interest in the 2D compass model is motivated by new opportunities
it provides for quantum computing \cite{Dou05}. This motivated also
plaquette orbital model (POM) introduced for a square lattice by Wenzel
and Janke \cite{Wen09} which exhibits orientational long-range order in
its classical version \cite{Bis10}. Here we will focus on the 1D quantum
version of the POM and investigate the nature of the ground state and of
low energy excitations. The purpose of this paper is to highlight the
importance of entangled states which lead to pronounced dimer
correlations in the 1D POM which consists of repeated interactions of
$x$ and $z$ pseudospin component along three bonds of a plaquette,
called for this reason also the $Cx$-$Cz$ model. As we show below, this
model has rather surprising properties which may be captured only in
analytic methods which go beyond standard mean-field (MF) approaches.

The paper is organized as follows: In Sec. \ref{sec:hami} we introduce
the $Cx$-$Cz$ Hamiltonian and derive its block-diagonal form making
use of its local symmetries. In Sec. \ref{sec:pert} we present a
perturbative approach to the model within its invariant subspaces up
to third order for the ground-state energies. The approximate solutions
of the model are presented in Sec. \ref{sec:mf} where we introduce a
single-dimer MF approach and more general two-dimer and three-dimer MF
approaches to show the ground state spin configuration in
different subspaces being the lowest excited states of the model. In
Sec. \ref{sec:ed} the exact diagonalization results are shown for the
maximal system size of $L=12$ dimers.
Finally, in Sec. \ref{sec:summa} we present the summary and main
conclusions. The paper is supplemented with two Appendices:
(i) Appendix \ref{sec:back} showing the additional details on spin
transformation used in Sec. \ref{sec:hami}, and
(ii) Appendix \ref{sec:dual} showing the duality between the interaction
and free terms in the block-diagonal $Cx$-$Cz$ Hamiltonian.

\section{Hamilonian and its symmetries}
\label{sec:hami}

\begin{figure}[t!]
\includegraphics[width=1\columnwidth]{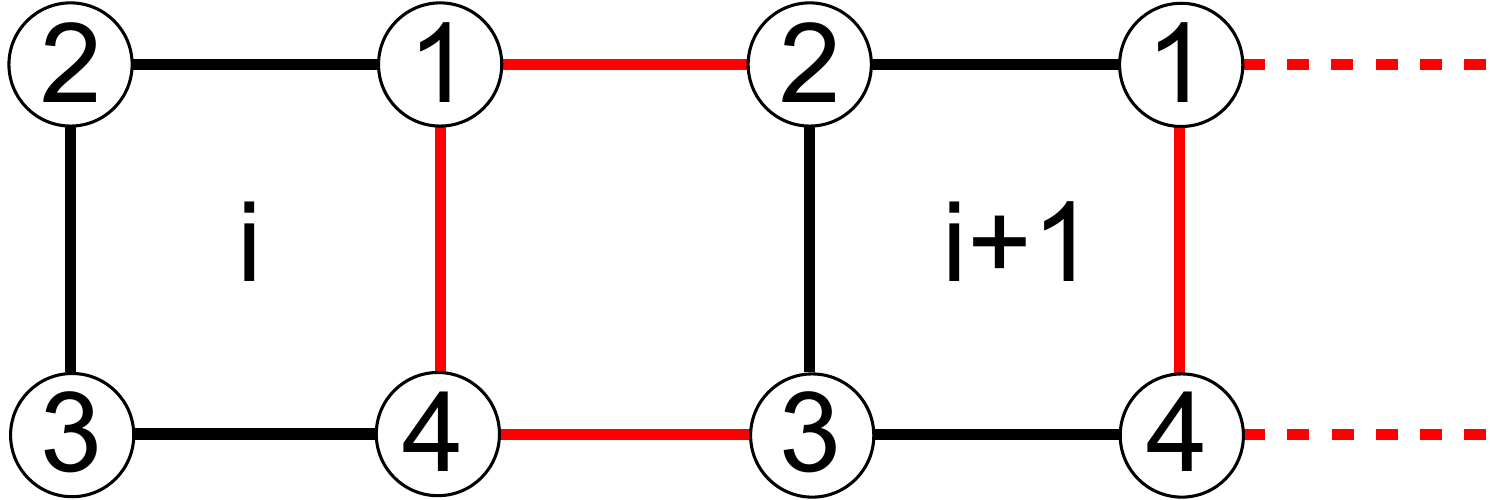}
\caption{(Color online)
Schematic view of the Hamiltonian of Eq. (\ref{eq:Ham}).
Black (red) lines stand for the\label{fig:plaqlad} $XX$ ($ZZ$)
bonds.}
\label{fig:pom}
\end{figure}

The Hamiltonian of the 1D POM ($Cx$-$Cz$ model) of $L$ sites can be
written as follows,
\begin{eqnarray}
{\cal H} & = & \sum_{i=1}^{L}\left\{
X_{i,1}X_{i,2}+X_{i,2}X_{i,3}+X_{i,3}X_{i,4}\right.\nonumber \\
&+& \left.Z_{i,1}Z_{i+1,2}+Z_{i,1}Z_{i,4}+Z_{i,4}Z_{i+1,3}\right\} ,
\label{eq:Ham}
\end{eqnarray}
where $X_{i,p}$ and $Z_{i,p}$ are the $x$ and $z$ Pauli matrices at
site $p$ of the plaquette $i$ --- see Fig. \ref{fig:plaqlad}. We assume
periodic boundary conditions (PBCs) of the form
$Z_{L+1,2}\equiv Z_{1,2}$ and $Z_{L+1,3}\equiv Z_{1,3}$. There are two
types of the symmetry operators specific to the model, namely:
\begin{eqnarray}
\label{piz}
P_{i}^{z}&=&Z_{i,1}Z_{i,2}Z_{i,3}Z_{i,4},  \\
\label{pix}
P_{i}^{x}&=&X_{i+1,2}X_{i,1}X_{i,4}X_{i+1,3}.
\end{eqnarray}
In what follows
we will make use of these symmetries to find a block-diagonal form
of the Hamiltonian ${\cal H}$ by two consecutive spin transformation.

\begin{figure}
\includegraphics[width=0.75\columnwidth]{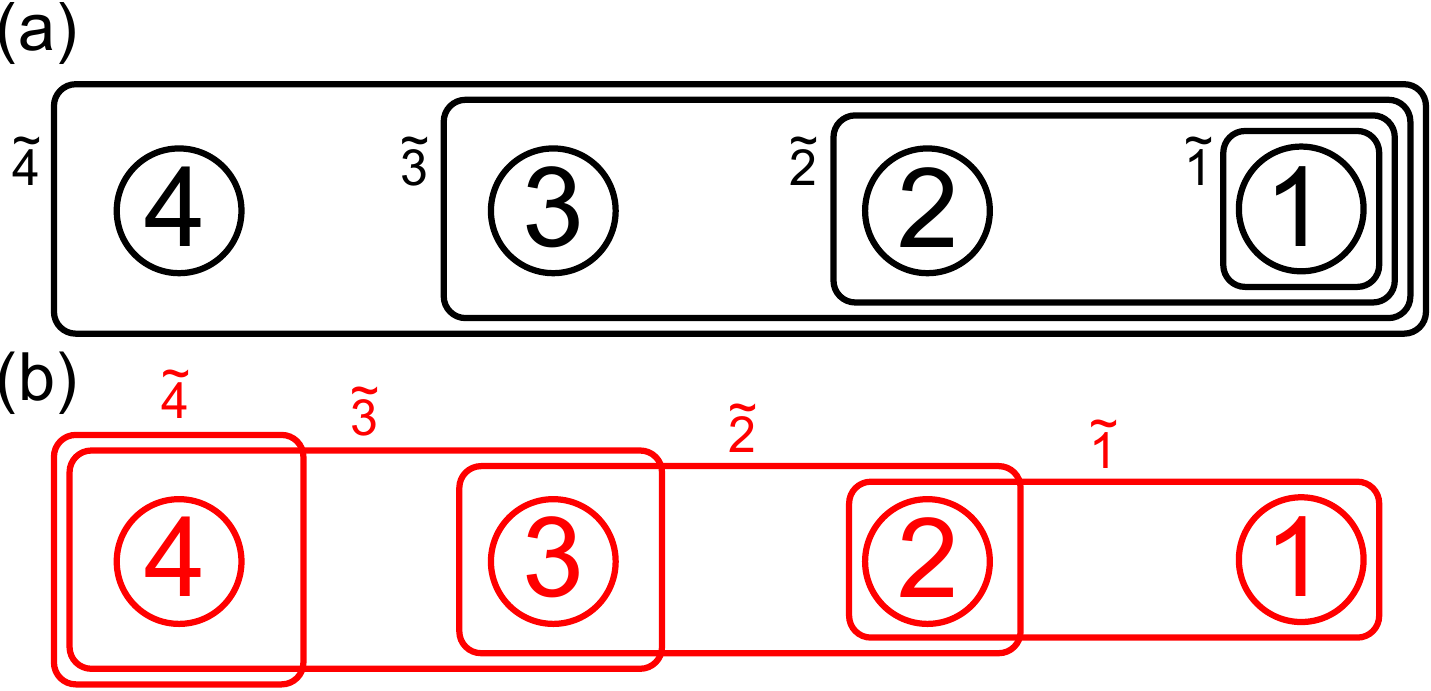}
\caption{(Color online)
Schematic view of the transformations used for the OPM:
(a) in Eqs. (\ref{eq:Xtilde}), and
(b) in Eq. (\ref{eq:Ztilde}).
Numbered circles symbolize otiginal Pauli matrices,
$\left\{ X_{i,1},X_{i,2},X_{i,3},X_{i,4}\right\}$ (black circles) or
$\left\{ Z_{i,1},Z_{i,2},Z_{i,3},Z_{i,4}\right\}$ (red circles).
The frames labeled with tilded numbers symbolize new Pauli matrices,
$\left\{\tilde{X}_{i,1},\tilde{X}_{i,2},\tilde{X}_{i,3},\tilde{X}_{i,4}\right\}$
(black frames), or
$\left\{\tilde{Z}_{i,1},\tilde{Z}_{i,2},\tilde{Z}_{i,3},\tilde{Z}_{i,4}\right\}$
(red frames).
\label{fig:trans}}
\end{figure}

The key observation for Pauli matrices defined on a product space
of a many-body system is that a product of $Z_{i,p}$ ($X_{i,p}$)
operators over any subset of the system is another $z$ ($x$) Pauli
operator. Of course, to transform {\it all} $Z_{i,p}$ ($X_{i,p}$)
operators into new ones one has to choose these subsets carefully
to keep track of the canonical commutation relations, saying that
$z$ and $x$ Pauli operators having the same site index anticommute
and otherwise commute. This we can assure by checking the intersections
of the subsets over which the products are taken; if the intersection
contains odd number of sites then the new $z$ and $x$ Pauli operators
will anticommute, in opposite case they will commute. One can easily
verify that these rules are satisfied by the transformation that we
use to take care of the $P_{i}^{z}$ symmetries of Hamiltonian
(\ref{piz}). The transformation is defined for each $X$-plaquette
separately as,
\begin{eqnarray}
X_{i,1} &=& \tilde{X}_{i,1},,            \nonumber\\
X_{i,2} &=& \tilde{X}_{i,1}\tilde{X}_{i,2},               \nonumber\\
X_{i,3} &=& \tilde{X}_{i,1}\tilde{X}_{i,2}\tilde{X}_{i,3},\nonumber\\
X_{i,4} &=& \tilde{X}_{i,1}\tilde{X}_{i,2}\tilde{X}_{i,3}\tilde{X}_{i,4},
\label{eq:Xtilde}
\end{eqnarray}
and
\begin{eqnarray}
Z_{i,1} & = & \tilde{Z}_{i,1}\tilde{Z}_{i,2},\nonumber \\
Z_{i,2} & = & \tilde{Z}_{i,2}\tilde{Z}_{i,3},\nonumber \\
Z_{i,3} & = & \tilde{Z}_{i,3}\tilde{Z}_{i,4},\nonumber \\
Z_{i,4} & = & \tilde{Z}_{i,4}.
\label{eq:Ztilde}
\end{eqnarray}
The operators $\tilde{X}{}_{i,p}$ and $\tilde{Z}_{i,p}$ are new
$x$ and $z$ Pauli matrices satisfying all the canonical commutation
relations and the tranformation is a bijection which means that the
inverse transformation exists --- its form can be easily guessed if
we notice that, e.g. $\tilde{X}_{i,2}=X_{i,2}X_{i,1}$ and
$\tilde{Z}_{i,3}=Z_{i,4}Z_{i,3}$.
This of course exploits the fact that any Pauli matrix squared gives
identity. The easiest way to verify that the transformations given by
Eqs. (\ref{eq:Xtilde}) and (\ref{eq:Ztilde}) really map Pauli operators
into other set of Pauli operators is by drawing --- see Fig.
\ref{fig:trans}.

\begin{figure}[t!]
\includegraphics[width=0.75\columnwidth]{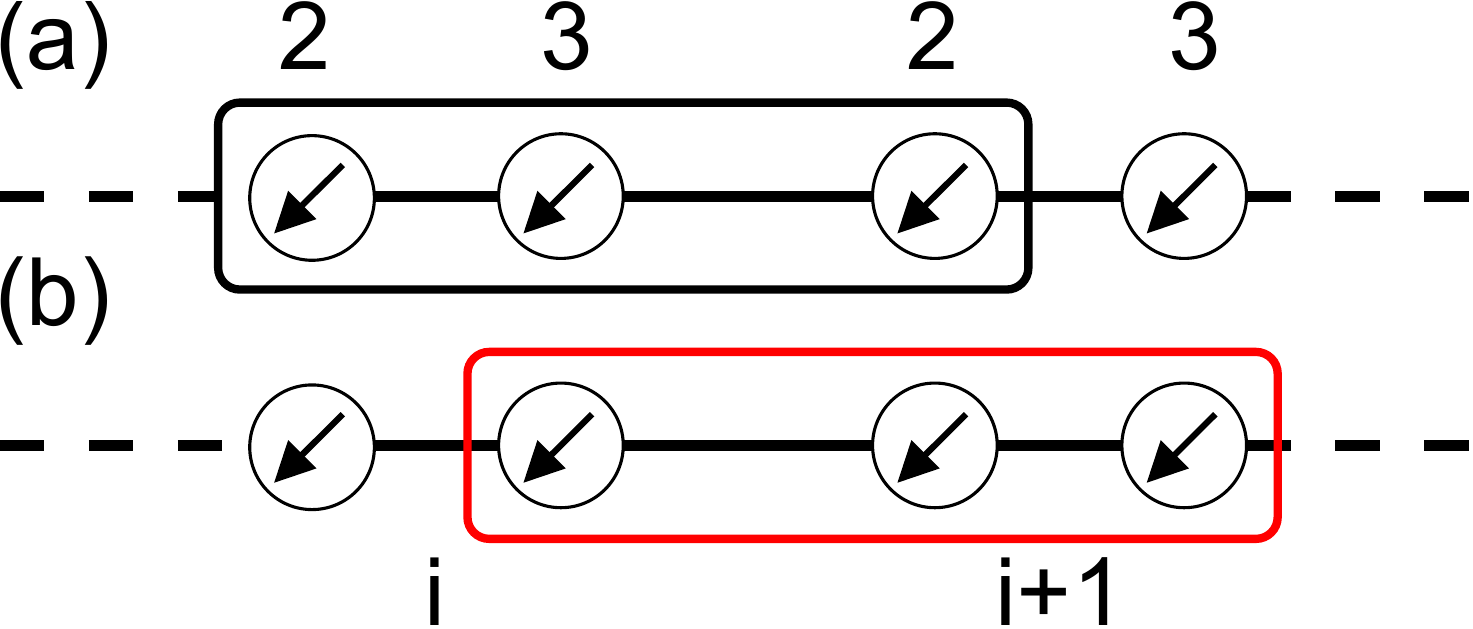}
\caption{(Color online) Schematic view of the interaction part of the
Hamiltonian Eq. (\ref{eq:Ham_sig}):
(a) the $x$ interactions (black frame), and
(b) the $z$ interactions (red frame).
The arrows represent the ground state configuration of the spins
$\sigma_{i,p}$ stabilized by the external field, under the assumption
that the interaction part is absent. }
\label{fig:ham_sig}
\end{figure}

It is straightforward to get the Hamiltonian in terms of tilde operators,
i.e.,
\begin{eqnarray}
{\cal H} &=& \sum_{i=1}^{L}\left\{
\tilde{X}_{i,2}+\tilde{X}_{i,3}+\tilde{X}_{i,4}
+r_{i}\tilde{Z}_{i,2}\tilde{Z}_{i+1,2}\tilde{Z}_{i+1,3}
\right.\nonumber\\
&+&\left.r_{i}\tilde{Z}_{i,2}\tilde{Z}_{i,4}
 + \tilde{Z}_{i,4}\tilde{Z}_{i+1,3}\tilde{Z}_{i+1,4}\right\},
\label{eq:Ham_tilde}
\end{eqnarray}
where $r_{i}=\pm1$ are the eigenvalues of the symmetry operator
$P_{i}^{z}=\tilde{Z}_{i,1}$,
which we are allowed to insert for ${\cal H}$ does not depend on
$\tilde{X}_{i,1}$. Consequently the $P_{i}^{x}$ symmetries transform as
\begin{equation}
P_{i}^{x}=X_{i+1,2}X_{i,1}X_{i,4}X_{i+1,3}=
\tilde{X}_{i,2}\tilde{X}_{i,3}\tilde{X}_{i,4}\tilde{X}_{i+1,3}.
\end{equation}
Now the hard part starts because this symmetry mixes the operators on
neighboring plaquettes. How to guess next spin transformation that will
make use of $P_{i}^x$ symmetries, provided that such a transformation
exists? We can try to demand that in terms of new Pauli operators the
symmetry transforms into a single Pauli operator, as it happened with
$P_{i}^{z}$, i.e., $P_{i}^{x}=X'_{i,4}$. This means that
$X'_{i,4}=\tilde{X}_{i,2}\tilde{X}_{i,3}\tilde{X}_{i,4}\tilde{X}_{i+1,3}$.
The form of the transformation (\ref{eq:Xtilde}) suggests that the
other $x$ operators can be constructed in the following way,
\begin{eqnarray}
X'_{i,2}&=& \tilde{X}_{i,2}\left(\tilde{X}_{i+1,3}\right),
\nonumber \\
X'_{i,3}&=& \tilde{X}_{i,2}\tilde{X}_{i,3}\left(\tilde{X}_{i+1,3}\right),
\nonumber \\
X'_{i,4}&=&
\tilde{X}_{i,2}\tilde{X}_{i,3}\tilde{X}_{i,4}\left(\tilde{X}_{i+1,3}\right),
\label{eq:Xprim_inv}
\end{eqnarray}
where the main difference with respect to Eq. (\ref{eq:Xtilde}) is that
we keep the contribution from the neighboring plaquette (in bracket) for
every $X'_{i,p}$. By analogy to the transformation (\ref{eq:Ztilde}) we
can also guess the form of the new $z$ operators,
\begin{eqnarray}
Z'_{i,2} &=& \tilde{Z}_{i,2}\tilde{Z}_{i,3}\left(\tilde{Z}_{i-1,2}\right),
\nonumber \\
Z'_{i,3} &=& \tilde{Z}_{i,3}\tilde{Z}_{i,4}\left(\tilde{Z}_{i-1,2}\right),
\nonumber \\
Z'_{i,4} &=& \tilde{Z}_{i,4}.
\label{eq:Zprim_inv}
\end{eqnarray}
Again the difference is in terms in brackets coming from the neighboring
plaquette - these were involved in Eq. (\ref{eq:Zprim_inv}) in such a
way that the canonical commutation relations between primed Pauli
operators are satisfied. Now to get the Hamiltonian in terms of new,
primed operators we need to inverse the above transformations. This
can be done in straightforward fashion and we arrive at,
\begin{eqnarray}
\tilde{X}_{i,2} & \!=\! & X'_{i,2}\left(\tilde{X}_{i+1,3}\right)
= X'_{i,2}\left(X'_{i+1,2}X'_{i+1,3}\right),\nonumber \\
\tilde{X}_{i,3} & \!=\! & X'_{i,2}X'_{i,3},\nonumber \\
\tilde{X}_{i,4} & \!=\! & X'_{i,3}X'_{i,4},\label{eq:Xprim}
\end{eqnarray}
and
\begin{eqnarray}
\tilde{Z}_{i,2} & \!=\! & Z'_{i,2}Z'_{i,3}Z'_{i,4},\nonumber \\
\tilde{Z}_{i,3} & \!=\! & Z'_{i,3}Z'_{i,4}\!\left(\tilde{Z}_{i-1,2}\right)
\!=\! Z'_{i,3}Z'_{i,4}\left(Z'_{i-1,2}Z'_{i-1,3}Z'_{i-1,4}\right),
\nonumber \\
\tilde{Z}_{i,4} & \!=\! & Z'_{i,4}.\label{eq:Zprim}
\end{eqnarray}
Quite miraculously these rather complicated formulas inserted into
Hamiltonian (\ref{eq:Ham_tilde}) give a rather simple structure of
the block-diagonal Hamiltonian,
\begin{eqnarray}
{\cal H} & = & \sum_{i=1}^{L}\left\{
s_{i}X'_{i,3}+X'_{i,2}X'_{i,3}+X'_{i,2}X'_{i+1,2}X'_{i+1,3}\right.
\nonumber \\
& + & \left.r_{i}Z'_{i+1,2}+r_{i}Z'_{i,2}Z'_{i,3}
  +Z'_{i+1,3}Z'_{i,2}Z'_{i,3}\right\} ,
\label{eq:Ham_prim}
\end{eqnarray}
where half of the initial spins are replaced by the quantum numbers
$r_{i},s_{i}=\pm1$ being the eigenvalues of the symmetry operators
$P_{i}^{z}$ and $P_{i}^{x}$. Thus a spin model on a ladder show
in Fig. \ref{fig:plaqlad} has become a model of a dimerized chain
with two spins per unit cell, namely $Z'_{i,2}$ and $Z'_{i,3}$.
Note that unlike in case of the 2D quantum compass model, where the
similar spin transformations were used to obtain reduced Hamiltonian
\cite{Brz10}, here the PBCs do not yield any non-local operators
in ${\cal H}$ of Eq. (\ref{eq:Ham_prim}). Here the PBCs assumed
for the initial spins become PBCs for both tilde operators of Eq.
(\ref{eq:Ham_tilde}) and primed ones of Eq. (\ref{eq:Ham_prim}).
Before discussing the reduced Hamiltonian in more details let us end
this Section by a one more (simple!) spin transformation that puts
${\cal H}$ in more symmetric and convenient form, namely
\begin{eqnarray}
\sigma_{i,3}^{x} & = & s_{i}X'_{i,3},\nonumber\\
\sigma_{i,2}^{x} & = & X'_{i,2}X'_{i,3},\nonumber \\
\sigma_{i,3}^{z} & = & r_{i}Z'_{i,2}Z'_{i,3},\nonumber \\
\sigma_{i,2}^{z} & = & r_{i-1}Z'_{i,2},
\label{eq:sig}
\end{eqnarray}
which finally gives,
\begin{eqnarray}
{\cal H} & = & \sum_{i=1}^{L}\left\{ \left(\sigma_{i,2}^{z}
+\sigma_{i,3}^{z}\right)+\left(\sigma_{i,2}^{x}
+\sigma_{i,3}^{x}\right)\right.\nonumber \\
& + &\left.r_{i}\sigma_{i-1,3}^z\left(\sigma_{i,2}^z\sigma_{i,3}^{z}\right)
+s_{i}\left(\sigma_{i,2}^{x}\sigma_{i,3}^{x}\right)\sigma_{i+1,2}^{x}\right\} .
\label{eq:Ham_sig}
\end{eqnarray}
This expression means that all the $\sigma_{i,p}$ spins are coupled to
an external magnetic field applied along direction $x+z$ and interact
by a three-spin interaction depicted in Fig. \ref{fig:ham_sig}, with
signs given by the $r_{i}$ and $s_{i}$ quantum numbers. The structure
of the interaction is such that we can consider the system as a set
of interacting dimers labeled by $i$ consisting of spins $\sigma_{i,2}$
and $\sigma_{i,3}$. In the Appendix \ref{sec:back}
we show the relation between $\sigma_{i,p}^{x,z}$ Pauli operators
and the original ones, $X_{i,p}$ and $Z_{i,p}$, of Eq. (\ref{eq:Ham}).

Finally, it is worth to mention that the structure of the free and
interaction terms in the Hamiltonian (\ref{eq:Ham_sig}) is strongly
related, i.e., we can find a basis where the linear terms become cubic
and vice-versa. As there are twice as many linear terms as the cubic
ones it is not possible to obtain a one-to-one correspondence between
the free and interacting part of the Hamiltonian --- in the Appendix
\ref{sec:dual} we give the additional interaction terms that should be
added to obtain such duality as well as the form of the duality spin
transformation.

\section{Perturbative treatment\label{sec:pert}}

The first question we may ask seeing the reduced Hamiltonian
(\ref{eq:Ham_sig}) of the 1D POM is in which subspace labeled by the
$r_{i}$ and $s_{i}$ quantum numbers the ground state can be found.
This can be easily answered by a perturbative expansion where the
unperturbed Hamiltonian ${\cal H}_0$ is the noninteracting part of
${\cal H}$, i.e.,
\begin{equation}
{\cal H}_{0} = \sum_{i=1}^{L}\left(\sigma_{i,2}^{z}
+\sigma_{i,2}^{x}+\sigma_{i,3}^{z}+\sigma_{i,3}^{x}\right),
\label{eq:H0}
\end{equation}
and the perturbation ${\cal V}$ is given by the three-spin terms,
\begin{equation}
{\cal V} = \sum_{i=1}^{L}\!\left\{ r_{i}\sigma_{i-1,3}^{z}
\!\left(\sigma_{i,2}^{z}\sigma_{i,3}^{z}\right)
\!+\! s_{i}\!\left(\sigma_{i,2}^{x}\sigma_{i,3}^{x}\right)\!
\sigma_{i+1,2}^{x}\right\}.
\label{eq:V}
\end{equation}

The ground state of $\left|0\right\rangle $ of ${\cal H}_{0}$ is easy
to infer, the spins order as in Fig. \ref{fig:ham_sig} with the ground
state energy per dimer equal to
\begin{equation}
E_{0}^{(0)}=-2\sqrt{2}\simeq -0.282.
\end{equation}
Hamiltonian
${\cal H}_{0}$ has a big energy gap of $\varepsilon_{1}=2\sqrt{2}$
which makes the expansion justified although formally there is no
small parameter in ${\cal V}.$ The first order correction to the
ground state energy is just the average of ${\cal V}$ in the state
$\left|0\right\rangle$
which is simple to calculate as we deal with a simple product state.
Thus we get a first order correction,
\begin{equation}
E_{0}^{(1)}=-\frac{1}{2\sqrt{2}L}\sum_{i=1}^{L}\left(r_{i}+s_{i}\right),
\label{eq:1st_order}
\end{equation}
as a linear function of the quantum numbers $r_{i}$ and $s_{i}$.
Now it is easy to see that the ground state of the model is in the
subspace with $r_{i}=s_{i}=1$ for all $i$. This result also suggests
that the lowest excited state of the model is the ground state from
the subspace with one $r_{i}$ or $s_{i}$ flipped --- such excitation
costs the energy of $1/\sqrt{8}\approx0.353$ in the leading order
while the excitation within the lowest subspace costs the energy of
$2\sqrt{2}\approx2.828$ in the leading order.

As the first order correction cannot be regarded as small we now proceed
to the higher orders. The second order correction has a form of,
\begin{equation}
E_{0}^{(2)}=-\frac{1}{L}\sum_{n\not=0}\frac{1}{\varepsilon_{n}}
\left\langle 0\right|{\cal V}\left|n\right\rangle
\left\langle n\right|{\cal V}\left|0\right\rangle,
\end{equation}
where $\varepsilon_{n}=E_{n}^{(0)}-E_{0}^{(0)}$ is the excitation
energy of the $n$-th excited state of ${\cal H}_{0}$. After a moderate
analytical effort we can get a correction,
\begin{equation}
E_{0}^{(2)}=\frac{-1}{2^{4}3\sqrt{2}}
\left\langle
6r_{i}r_{i+1}\!+\!6s_{i}s_{i+1}\!-\!9s_{i}r_{i+1}\!-\!9r_{i}s_{i}
\!+\!29\right\rangle,
\label{eq:2nd_order}
\end{equation}
where the interaction terms between the classical spins are present and
we take a contribution for the representative sites $i$ and $(i+1)$.
The value of $E_{0}^{(2)}$ in the ground subspace is
$E_{0}^{(2)}\approx-0.339$, the total ground state energy up to the
second order is equal to
\begin{equation}
E_{0}^{(0)}+E_{0}^{(1)}+E_{0}^{(2)}\simeq
- 2.828 - 0.707 - 0.339=-3.874\,.
\end{equation}
Such a value is problematic for, as we will see in the next Section,
the extrapolated ground state energy from the exact diagonalization is
equal to $E_{0}^{\rm ED}\simeq-3.7897$ which is higher than what we
have obtained up to second order.

The above result and overall largeness of the second order correction
indicates that we should go to the third order to get the energy within
the physical range of values. The textbook expression for the third
order energy correction reads,
\begin{eqnarray}
E_{0}^{(3)}L & \!=\! & \sum_{n\not=m\not=0}
\frac{1}{\varepsilon_{n}\varepsilon_{m}}
\left\langle 0\right|{\cal V}\left|n\right\rangle\left\langle n\right|
{\cal V}\left|m\right\rangle
\left\langle m\right|{\cal V}\left|0\right\rangle
\nonumber \\
& \!+\! & \sum_{n\not=0}\frac{
 \left\langle n\right|{\cal V}\left|n\right\rangle
-\left\langle 0\right|{\cal V}\left|0\right\rangle }
{\varepsilon_{n}^{2}}\left\langle 0\right|{\cal V}\left|n\right\rangle
\left\langle n\right|{\cal V}\left|0\right\rangle .
\label{eq:3rd_order_gen}
\end{eqnarray}
This already requires a considerable effort to calculate as due to the
canted nature of the unperturbed ground state there are not many
overlaps that cancel in the above expression. Probably the simplest way
to calculate this correction is to span the Hilbert space of possible
excited states for a given dimer $i$ in ${\cal V}$, define the operators
in the product space and calculate the correction by a brute force.
Here we used Mathematica to do it and the Hilbert space was a product
space of $11$ with a dimension of $2^{11}$ and the $r_{i}$ and
$s_{i}$ quantum numbers were kept as variables. The results is,
\begin{eqnarray}
E_{0}^{(3)} & \!=\! & \frac{1}{2^{6}3\sqrt{2}}
\left\langle22\!\left(r_{i}\!+\! s_{i}\right)\!-\!3\!\left(r_{i}s_{i}r_{i+1}
\!+\! s_{i}r_{i+1}s_{i+1}\right)\right\rangle\nonumber \\
& \!+\! & \frac{1}{2^{13}3}
\left\langle-29\left(r_{i}\!+\! s_{i}\right)\!-\!11\left(r_{i}s_{i}r_{i+1}
\!+\! s_{i-1}r_{i}s_{i}\right)\right.\nonumber \\
& \!+\! & 36\left.\left(r_{i-1}r_{i}s_{i}\!+\! s_{i-1}r_{i}r_{i+1}
\!+\! s_{i-1}s_{i}r_{i+1}\!+\! r_{i}s_{i}s_{i+1}\right)\right.\nonumber \\
& \!-\! & \left.24\left(r_{i-1}r_{i}r_{i+1}
+ s_{i-1}s_{i}s_{i+1}\right)\right\rangle,
\label{eq:3rd_order}
\end{eqnarray}
where we take again an average contribution for the representative sites
$i$ and $(i+1)$. Here the first line is a leading term that originates
from the contributions where the two intermediate states are the same,
i.e., $n=m$ --- the second line of Eq. (\ref{eq:3rd_order_gen}).
The third order correction to the ground state is positive and equal
to $E_{0}^{(3)}\approx0.140$. Thus the ground state energy up to third
order is $E_{0}^{\rm pert}\approx-3.734$ which is now well within the
physical range given by the ED reported in Sec. \ref{sec:ed} --- the
energy difference between this result and $E_{0}^{\rm ED}$ is of the
order of $0.05$ so one can conclude that the third order expansion is
{\it almost} exact.

\begin{figure}[t!]
\includegraphics[width=1\columnwidth]{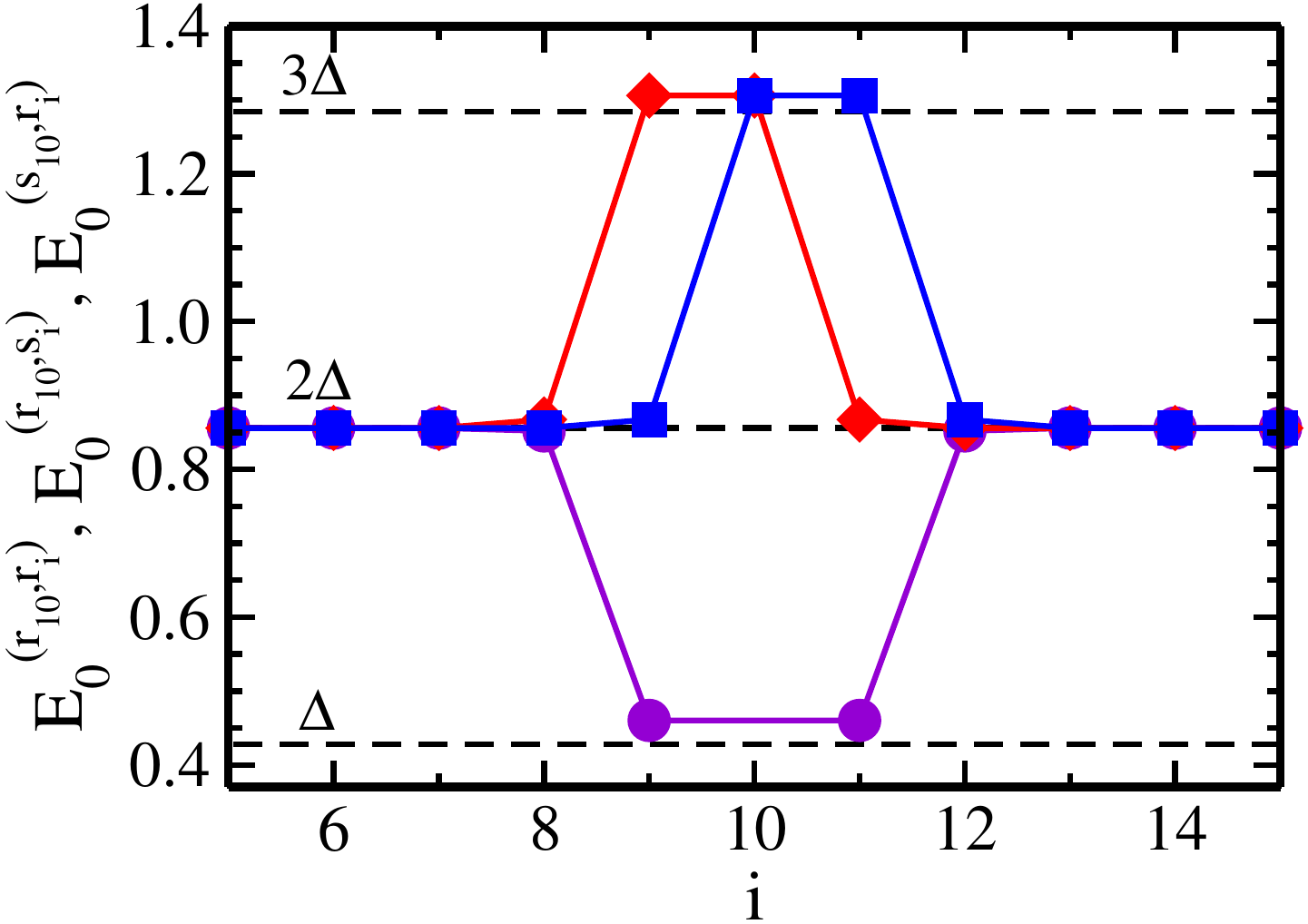}
\caption{(Color online) Ground state energies $E_{0}^{(p_{1},p_{2})}$
from the subspaces with two classical spins $p_{1,2}$ being excited;
dots --- $r_{10}$ and $r_{i}$ flipped, diamonds --- $r_{10}$ and
$s_{i}$ flipped and squares --- $s_{10}$ and $r_{i}$ flipped, as
functions of $i$. Dashed lines show the energies of single, double and
triple energy gap $\Delta$.  }
\label{fig:two_flips}
\end{figure}

Concerning the excitations, the energy gap given by the expansion
is $\Delta^{\rm pert}=0.428$ which is close to the ED value of the gap
$\Delta^{\rm ED}=0.437$, see below in Sec. \ref{sec:ed}. As stated
earlier, the first excited state is the ground state of the model in
the subspace with one $r_{i}$ or $s_{i}$ being flipped. Eq.
(\ref{eq:1st_order}) and the value of $\Delta^{\rm pert}$ suggests that
flipping two classical spins $r_{i}$ or $s_{i}$ should still cost less
energy than creating an excitation within the ground subspace. We may
expect that if the defects in the configuration of classical spins are
sufficiently far from each other then the excitation energy should be
$2\Delta^{\rm pert}$.

In Fig. \ref{fig:two_flips} we show the excitation energies for one
defect placed at site $i_{1}=10$ and second at any other site $i$ as a
function of $i$. As at every site we have both $r_{i}$ and $s_{i}$
there are four possibilities of creating such a pair of defects because
for each site we can flip $r_{i}$ or $s_{i}$. Due to the symmetry of
Eq. (\ref{eq:Ham_sig}) flipping two $r_{i}$'s is equivalent to flipping
two $s_{i}$'s. As we can see from Fig. \ref{fig:two_flips}, all the
excitation energies are close to $2\Delta$ when the defects are
separated by more than two sites --- this is additive regime governed
by the first order correction of Eq. (\ref{eq:1st_order}).

When the distance is smaller then we observe two different behaviors,
the gap for $r$-$r$ (or $s$-$s$) excitation is smaller than expected
and close to $\Delta$ and the gap for $s$-$r$ (or $r$-$s$) excitation
is bigger than expected and close to $3\Delta$. In this regime the
second and third order corrections are important. Such behavior means
that flipping classical spins of different flavors at neighboring or
the same sites is something that the system particulary dislikes.
On the other hand, if we choose
only one flavor to flip then Fig. \ref{fig:two_flips} suggests that
we could even flip {\it all} the $r_{i}$'s paying only one $\Delta$
of the excitation energy. The ED results show that this is not true
(the higher order correction are important in this case) however they
show that flipping all $r_{i}$'s still costs less energy than an
excitation within the ground subspace.

\begin{figure}[t!]
\includegraphics[width=0.75\columnwidth]{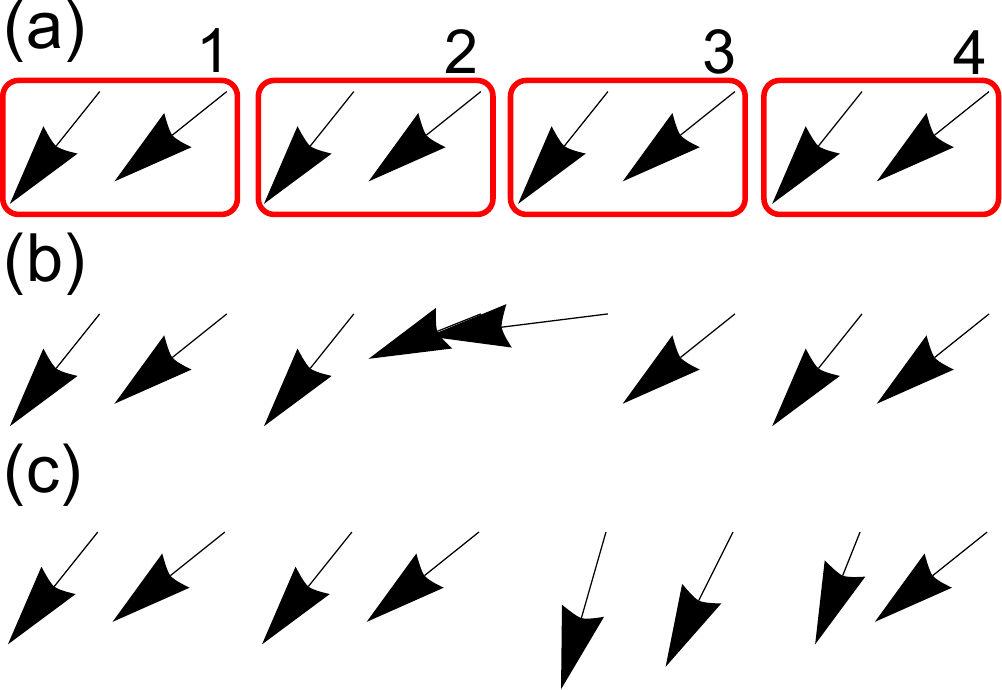}
\caption{(Color online)
Local spin averages $\left\langle \sigma_{2,i}^{x,z}\right\rangle$
and $\left\langle \sigma_{3,i}^{x,z}\right\rangle$ shown as arrows
in the:
(a) global ground state,
(b) first excited state with $r_{3}=-1$, and
(c) first excited state with $s_{3}=-1$.
The horizontal (vertical) components of the vectors (arrows)
correspond to the $x$($z$) components of the spins $\sigma_{p,i}$.
The frames indicate the dimers
$\left\{\sigma_{2,i},\sigma_{3,i}\right\}$ with $i=1,2,3,4$.
\label{fig:pert_arrows} }
\end{figure}

Finally, using the perturbation approach it is possible to look not
only at the energies in different subspaces but also at the ground state
spin configuration. It is quite simple to check that up to the first
order the local spin averages are given by the following formulas,
\begin{eqnarray}
\left\langle \sigma_{2,i}^{x}\right\rangle  & = &
-\frac{1}{\sqrt{2}}+\frac{2r_{i}-s_{i}-s_{i-1}}{4\sqrt{2}},\nonumber\\
\left\langle \sigma_{2,i}^{z}\right\rangle  & = &
-\frac{1}{\sqrt{2}}+\frac{s_{i}-2r_{i}}{4\sqrt{2}},\nonumber \\
\left\langle \sigma_{3,i}^{x}\right\rangle  & = &
-\frac{1}{\sqrt{2}}-\frac{s_{i}}{4\sqrt{2}},\nonumber \\
\left\langle \sigma_{3,i}^{z}\right\rangle  & = &
-\frac{1}{\sqrt{2}}+\frac{s_{i}-r_{i+1}}{4\sqrt{2}}.
\end{eqnarray}
In Fig. \ref{fig:pert_arrows} we show the above averages represented
by the arrows for four sites $i=1,2,3,4$ with PBCs for the global
ground state, shown in Fig. \ref{fig:pert_arrows}(a), and the lowest
excited states with $r_{3}=-1$ and $s_{3}=-1$, see Figs.
\ref{fig:pert_arrows}(b) and \ref{fig:pert_arrows}(c). In the ground
state we observe a two-sublattice order where the configuration of
neighboring spins differ by the interchange of the $x$ and $z$
component. In the excited states we observe distortion
of the spin order being different for a flip in $r$ and $s$ spins.
In the former case the $z$ components of the spins decrease when
approaching the site with defect and then grow again. In the latter
case the same happens to $x$ components so we can conclude that the
two excitations are complementary (this is also visible in Fig.
\ref{fig:two_flips}).

\section{Mean-field treatment}
\label{sec:mf}

\begin{figure}[t!]
\includegraphics[width=1\columnwidth]{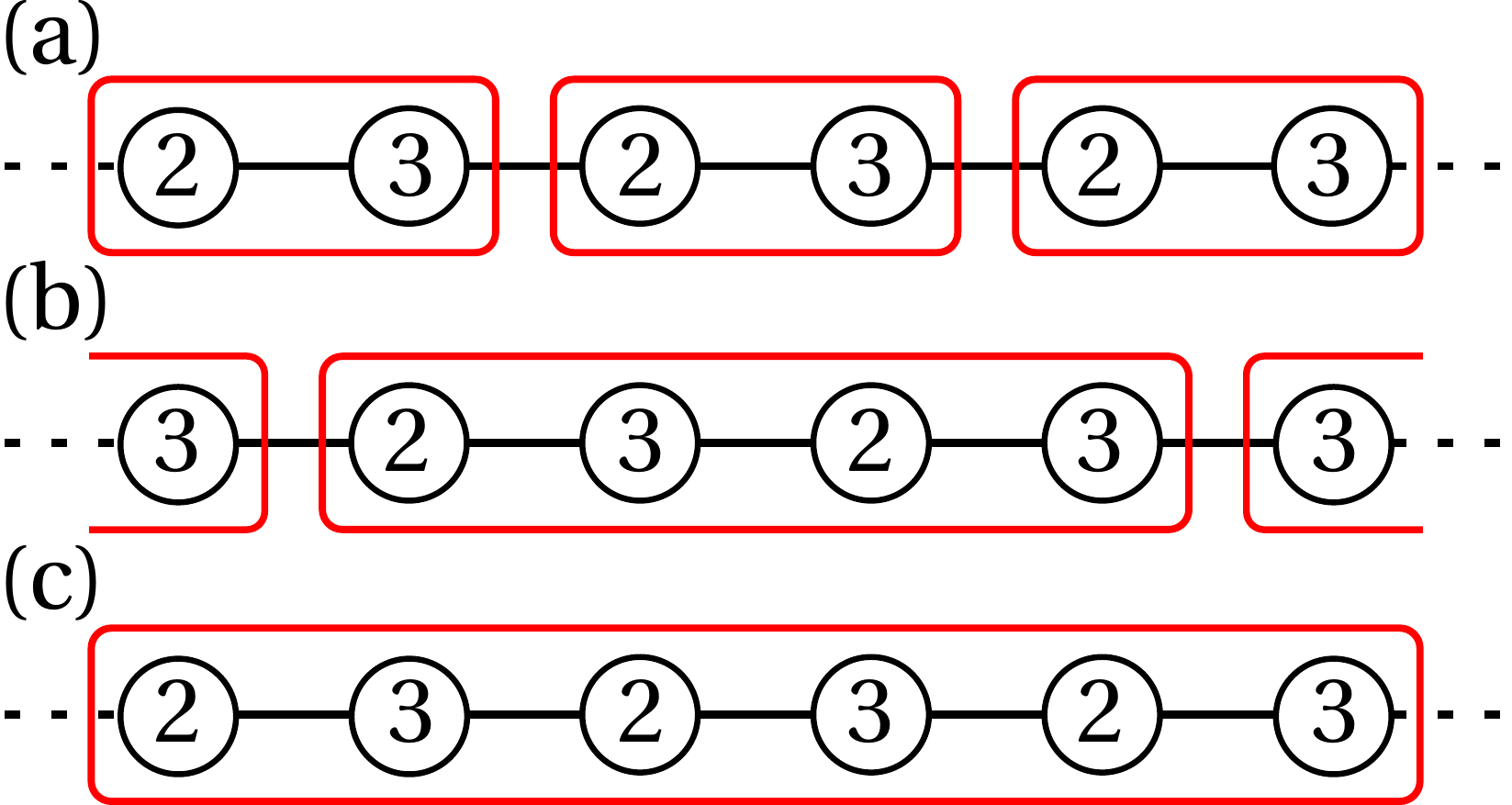}
\caption{(Color online) Schematic view of the dimer MF
decoupling in case of:
(a) one-dimer MF,
(b) two-dimer MF and
(c) three-dimer MF approximation.
The frames mark the cluster of dimers that are treated exactly.  }
\label{fig:mf}
\end{figure}

We have shown above that the excitation in the classical spins $r_{i}$
and $s_{i}$ are typically lower than a ``quantum'' excitation within
the ground subspace. Such excited states are on the other hand the
ground states of the reduced Hamiltonian (\ref{eq:Ham_sig}) in the
subspaces where some of the $r_{i}$'s or $s_{i}$'s are negative. This
suggests that these states can be well described within a nonuniform
MF approach carried out in any given subspace. The dimerized
form of the Hamiltonian (\ref{eq:Ham_sig}), see Fig. \ref{fig:ham_sig},
suggests a MF approach where a main building block is a dimer.
Thus if we think of a one-dimer approach we need to divide a system
into clusters containing one dimer each (see Fig. \ref{fig:mf}(a)) or
containing two, three dimers or more dimers [see Figs. \ref{fig:mf}(b)
and \ref{fig:mf}(c)] if we think of a more general approach.

Clusterization means that the interactions within a cluster are treated
exactly but different clusters interact only by MFs. This involves a
standard decoupling of the interaction terms in the Hamiltonian
(\ref{eq:Ham_sig}) assuming that the correlations between the clusters
are not strong, i.e.,
\begin{eqnarray}
\left(\sigma_{i,2}^{x}\sigma_{i,3}^{x}\right)\sigma_{i+1,2}^{x}
& \simeq & \left\langle \sigma_{i,2}^{x}\sigma_{i,3}^{x}\right\rangle
\sigma_{i+1,2}^{x}+\sigma_{i,2}^{x}\sigma_{i,3}^{x}
\left\langle \sigma_{i+1,2}^{x}\right\rangle \nonumber \\
& - & \left\langle \sigma_{i,2}^{x}\sigma_{i,3}^{x}\right\rangle
\left\langle \sigma_{i+1,2}^{x}\right\rangle ,\label{eq:x_dec}\\
\sigma_{i-1,3}^{z}\left(\sigma_{i,2}^{z}\sigma_{i,3}^{z}\right)
& \simeq & \sigma_{i-1,3}^{z}\left\langle\sigma_{i,2}^{z}\sigma_{i,3}^{z}
\right\rangle +\left\langle \sigma_{i-1,3}^{z}\right\rangle
\sigma_{i,2}^{z}\sigma_{i,3}^{z}\nonumber \\
 & - & \left\langle \sigma_{i-1,3}^{z}\right\rangle \left\langle
\sigma_{i,2}^{z}\sigma_{i,3}^{z}\right\rangle .
\label{eq:z_dec}
\end{eqnarray}
From this decoupling we have four independent MFs per dimer,
i.e., $\left\langle\sigma_{i,2}^{x}\sigma_{i,3}^{x}\right\rangle$,
$\left\langle \sigma_{i,2}^{z}\sigma_{i,3}^{z}\right\rangle $,
$\left\langle \sigma_{i,2}^{x}\right\rangle$, and
$\left\langle \sigma_{i,3}^{z}\right\rangle$.
In the case when the configuration of the classical spins is uniform
and the Hamiltonian (\ref{eq:Ham_sig}) is translationally invariant we
can safely assume that the above MFs do not depend on $i$ and
the self-consistency equations can be solved for any system size $L$.
This however is not the case in the excited subspaces that we are
interested in. Thus, typically, we need to work with a finite system
--- here we have taken $L=100$.

The self-consistency equations can be solved iteratively in each case,
i.e., we set some random initial values of the MFs, then we
diagonalize all the clusters and calculate new values of the MFs.
The procedure is repeated until the desired convergence of the MFs
is reached. In the majority cases this happens very quickly ---
after less than 100 iterations the old and the new value of each MF
field does not differ by more than $10^{-14}$. This however does not
refer to the subspaces with large areas being {\it fully defected},
i.e., for many neighboring sites $i$ we have $r_{i}=s_{i}=-1$. For
instance, if we set {\it all} classical spins as $-1$ then the two
interesting things happen within the MF approach:
(i) within the uniform approach no convergence is reached and
(ii) within a non-uniform approach we get a disordered configuration
which depends on the initial values of the MFs.
As we will see in the next Section such configuration is cured by the
quantum fluctuations and the true ground state has a two-sublattice
long-range order but with ordered moments that are strongly reduced
with respect to the ground state configuration and, as it will be shown
in Sec. \ref{sec:ed}, strongly enhanced entanglement between the
neighboring dimers. Finally, in order to check if the MF approximation
is justified we can extend it in a perturbative manner. What is omitted
in the MF approach are the correlation, so the full Hamiltonian
${\cal H}$ can be recovered from the MF one, ${\cal H}_{\rm MF}$,
by adding the missing many-body term of the form,
\begin{eqnarray}
{\cal V}_{\rm corr} & \!=\! & \sum_{i=1}^{L}\left\{
s_{i}\left(\sigma_{i,2}^{x}\sigma_{i,3}^{x}\!-\!\left\langle
\sigma_{i,2}^{x}\sigma_{i,3}^{x}\right\rangle\right)
\left(\sigma_{i+1,2}^{x}\!-\!\left\langle \sigma_{i+1,2}^{x}
\right\rangle\right)\right. \nonumber \\
& \!+\! & \left.r_{i}\left(\sigma_{i,2}^{z}\sigma_{i,3}^{z}
\!-\!\left\langle\sigma_{i,2}^{z}\sigma_{i,3}^{z}\right\rangle\right)
\left(\sigma_{i-1,3}^{z}\!-\!\left\langle \sigma_{i-1,3}^{z}
\right\rangle\right)\right\}.
\label{eq:V_cor}
\end{eqnarray}

Now we can write that
\begin{equation}
{\cal H}={\cal H}_{\rm MF}+{\cal V}_{\rm corr},
\end{equation}
and treat the many-body term as a perturbation. Due to the
self-consistency equations the first order correction to the energy
vanishes. The calculation of the second order correction is elementary
and requires the values of the MFs obtained earlier. It is significant
that the value of this second order correction is less than $2\%$ of the
MF energy in case of the ground state whereas it is almost $10\%$ of the
MF energy for the fully defected subspace. This means that the simple
MF approach works extremely well when no frustration is present and
much worse when its magnitude is maximal.

\begin{table}[b!]
\begin{center}%
\caption{Summary of the ground state energies $E_{0}$ (per dimer)
and the gap $\Delta$ obtained in the perturbation theory
(up to third order) and within the MF approaches compared with
the exact diagonalization results.}
\begin{ruledtabular}
\begin{tabular}{cccccc}
&       approach        &   $E_{0}$  &  $\Delta$ & \cr
\colrule
&  perturbation theory  &  $-3.734$  &  $0.428$  & \cr
&       1-dimer MF      &  $-3.6501$ &  $0.4134$ & \cr
&       2-dimer MF      &  $-3.7192$ &  $0.4520$ & \cr
&       3-dimer MF      &  $-3.7428$ &  $0.4592$ & \cr
& 1-dimer MF+correction &  $-3.7022$ &  $\dots$  & \cr
& exact diagonalization & $-3.789718$& $0.437271$& \cr
\end{tabular}
\end{ruledtabular}
\end{center}
\label{tab:energies}
\end{table}

To summarize these energetic considerations we present the ground state
energies obtained in perturbation theory and within the MF approach
using a 1-dimer (with and without a second order correction), 2-dimer,
and 3-dimer ansatz, respectively, compared to the value obtained by the
exact diagonalization in Table I. This latter energy we believe to be
the accurate one up to 6-digit precision (see Sec. \ref{sec:ed}). As we
can see, including one more dimer to the single-dimer MF improves the
energy by roughly $0.06$ whereas the second correction gives $0.05$.
On the other hand, adding another dimer lowers the energy to the value
which is very close to the estimated value, the difference is of the
order of $1\%$ only.

The excitation gap $\Delta$ requires good accuracy for both the
ground state energy and the ground state in the subspace of the first
excitation. Here the perturbation theory works somewhat better than the
MF ans\"atze, see Table \ref{tab:energies}. The method we developed for
the 1-dimer MF with a correction term \eqref{eq:V_cor} is reliable when
the calculated state is unform, so it is not used to estimate the value
of $\Delta$.

\begin{figure}[t!]
\includegraphics[width=1\columnwidth]{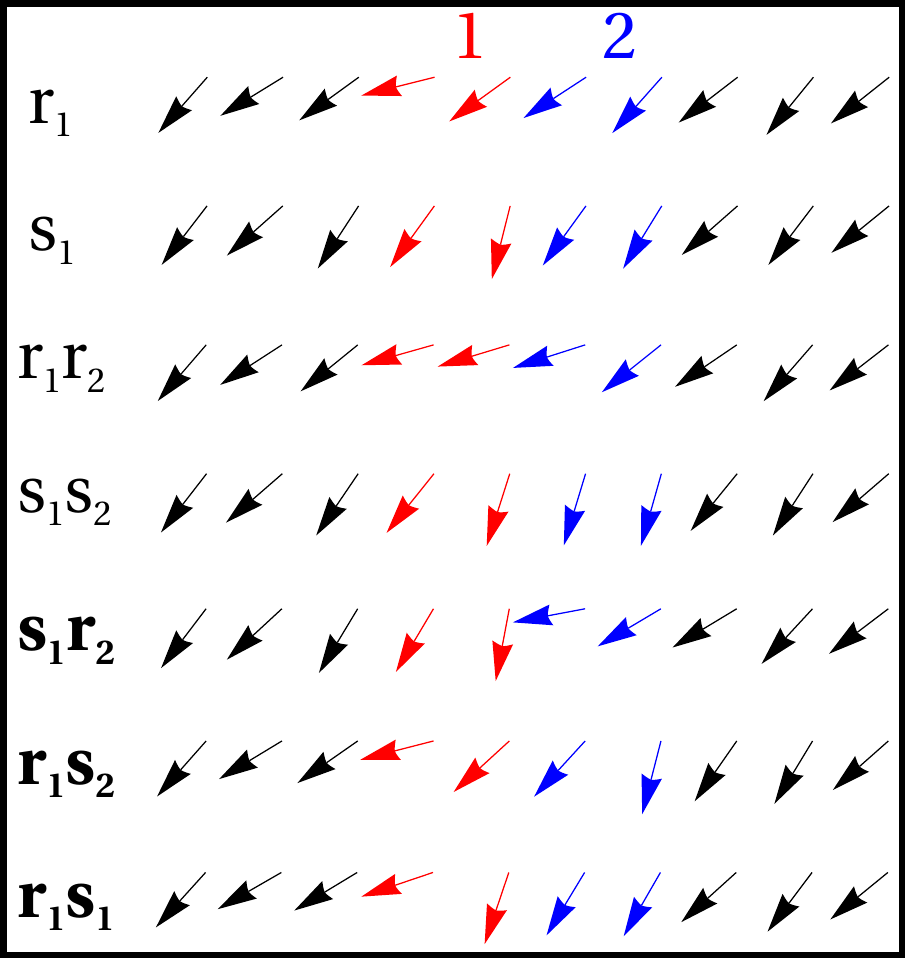}
\caption{(Color online)
Local spin averages $\left\langle \sigma_{2,i}^{x,z}\right\rangle $
and $\left\langle \sigma_{3,i}^{x,z}\right\rangle $ shown as arrows in
different excited states with such classical spins flipped as indicated
on the left. The horizontal (vertical) components of the vectors
(arrows) correspond to the $x$($z$) components of the spins
$\sigma_{p,i}$. The dimers $i=1,2$ are marked with different colors.
The classical spins that are flipped are denoted on the left and the
three last configurations
with high excitation energies are marked with bold face.  }
\label{fig:flips}
\end{figure}

\begin{figure}[t!]
\includegraphics[width=1\columnwidth]{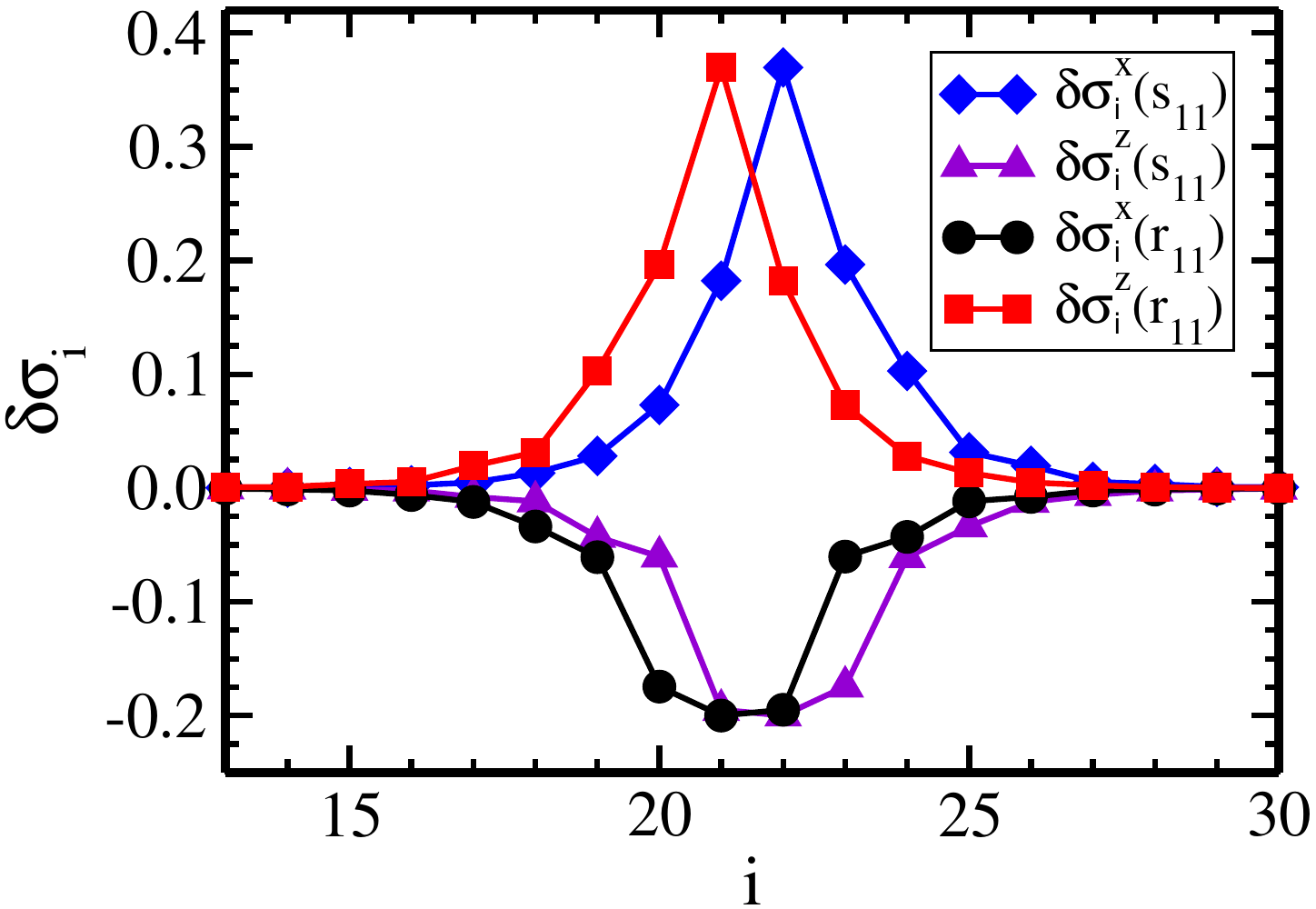}
\caption{(Color online) Differences in local spin averages in the
excited states with $r_{11}=-1$ or $s_{11}=-1$ with respect to their
ground state values,
$\delta\sigma_{i}^{x,z}(r_{11})$ and $\delta\sigma_{i}^{x,z}(s_{11})$.
Here we label the spins $\sigma_{2,i}$ and $\sigma_{3,i}$ by a single
index $i$ so that $r_{11}$ and $s_{11}$ refer to $11$th dimer
or spins $\sigma_{21}$ and $\sigma_{22}$.\label{fig:diffs_rs} }
\end{figure}

The MF spin configurations in the lowest excited states are shown in
Fig. \ref{fig:flips}. First two lines show the effect of a single
defect in $r_{i}$ and $s_{i}$ spins, respectively. These configurations
are qualitatively similar to the perturbative ones shown in Fig.
\ref{fig:pert_arrows} but the range of the distortion caused by the
defect is longer than before. In Fig. \ref{fig:diffs_rs} we show the
differences between these configurations and the ground state one ---
the distortion dies off at the distance of roughly $6$ spins
($3$ dimers). As shown in the plot of Fig. \ref{fig:two_flips} the
configurations with two defects have a doubled excitation energy with
respect to the ones with single defect when the defects are far apart.
In the next two lines of Fig. \ref{fig:flips} we show the configurations
with two defect only in $r_{i}$'s and only in $s_{i}$'s being next to
each other. As we know from Fig. \ref{fig:two_flips} such defects give
a sub-additive energy close to a single energy gap. Fig \ref{fig:flips}
shows that these configurations are indeed very similar to the ones
with single defects --- the range of distortion is longer but the
distortion itself is smoother. Finally, in the last three lines of Fig.
\ref{fig:flips} we show the two-defect cases when the excitation is
increased above the additive level. These configurations are
characterized by a rather severe spin distortion at the defects dimers
which is related with the local frustration caused by the defects and
is consistent with the increase of excitation energy.

According to Eq. (\ref{eq:sigxz_XZ}) it is possible to uniquely relate
the direction of the arrows shown in Fig. \ref{fig:flips} with the
values of bond operators of original ladder Hamiltonian of Eq.
(\ref{eq:Ham}). $\sigma_{i,2}^{x}$ and $\sigma_{i,2}^{z}$ operators are
the horizontal bonds within the $x$ and $z$ plaquettes respectively.
Similarly, $\sigma_{i,3}^{x}$ and $\sigma_{i,3}^{z}$ are the vertical
bonds within the $x$ and $z$ plaquettes. When an arrow points in
the direction $-\left(x+z\right)$, as it happens in the ground state,
it means that both $x$ and $z$ are locally satisfied. An arrow being
more horizontal than the others indicates that locally the $x$ bonds
are favored on expense of the $z$ ones. Analogically, a vertical tilt
means that the $z$ bonds are favored.

As we can see from Fig. \ref{fig:flips} excitation in $r_{i}$, which
is related with the $z$ interaction term in the reduced Hamiltonian
Eq. (\ref{eq:Ham_sig}), transfers the energy from $z$ to $x$ bonds
around site $i$. Excitation in $s_{i}$ has an inverse results. This we
can understand very easily. Assume that in Eq. (\ref{eq:Ham_sig}) we
have only the part with $z$ Pauli operators. When all $r_{i}$ are
positive than it is easy to check that in the ground state all spins
will be pointing down and every term in the Hamiltonian will give a
contribution $-1$ to the ground state energy. However if for one site
$i$ we set $r_i=-1$ then the frustration occurs because the linear part
of the Hamiltonian still wants all spins to point down whereas the
cubic part for site $i$ has now an opposite sign and for such spin
configuration gives a positive contribution to the energy. Thus the
cubic term is frustrated with the linear terms. When the defect is only
in the $r_{i}$ configuration then this frustration can be avoided by
adjusting the spin configuration more to the $x$ part of the
Hamiltonian and this exactly gives the horizontal tilt that we can see
in the first line of Fig. \ref{fig:flips}. On the other hand when both
$r_i$ and $s_i$ are locally negative then the frustration cannot be
avoided and a severe distortion in the spin configuration occurs as
shown in Fig. \ref{fig:flips}. In Sec. \ref{sec:ed} we will demonstrate
that such frustration can also lead to local disorder with increased
quantum entanglement.

\section{Exact diagonalization treatment}
\label{sec:ed}

\begin{figure}[t!]
\includegraphics[width=0.75\columnwidth]{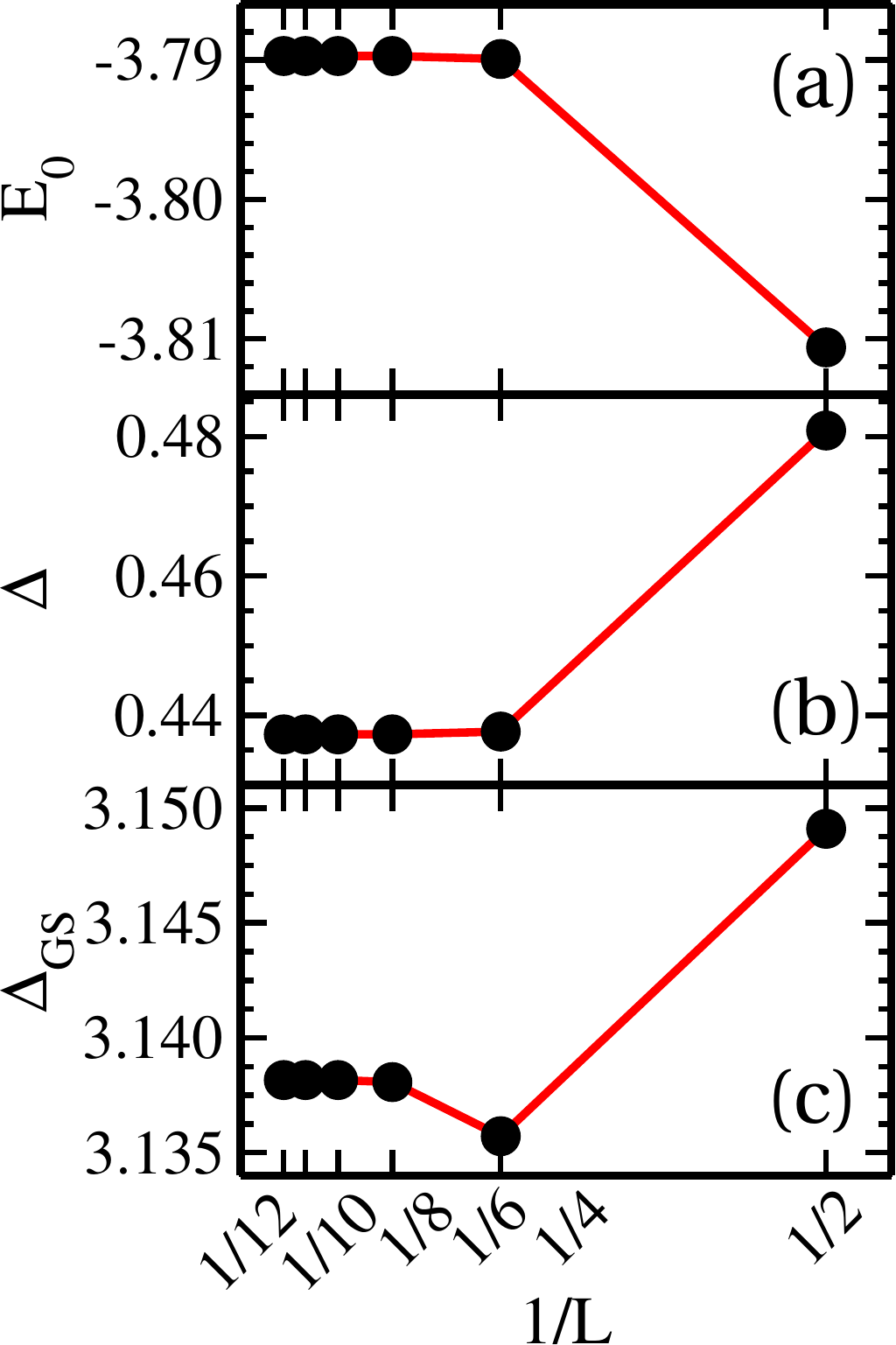}
\caption{(Color online) Finite size scaling obtain for the POM:
(a) ground state energy per dimer $E_{0}$,
(b) energy gap $\Delta$, and
(c) energy gap in the ground subspace $\Delta_{\rm GS}$. }
\label{fig:extr_ene}
\end{figure}

Exact diagonalization was carried out using Lanczos algorithm for the
system sizes up to $L=12$ for even $L$ and PBCs. In Fig.
\ref{fig:extr_ene}(a) we show finite size scaling of the ground state
energies per dimer as function of $1/L$. Quite remarkably the energy
saturates very quickly so the last four values are the same up to seven
digits. A similar behavior is observed for the energy gap $\Delta$, see
Fig. \ref{fig:extr_ene}(b). Thus we can conclude that the values of the
ground state energy (per dimer) and the energy gap obtained for $L=12$
are good approximations for the infinite system, these are:
\begin{eqnarray}
\label{egs}
 E_{0}^{\rm ED}&=&-3.789718,\\
\label{delta}
 \Delta^{\rm ED}&=& 0.437271.
\end{eqnarray}

Although a gap within a ground-subpace $\Delta_{\rm GS}$, shown in Fig.
\ref{fig:extr_ene}(c), exhibits less regular scaling behavior but again
a quick saturation is observed for the last three points so we can
treat the last point as the thermodynamic limit, thus we have found
that
\begin{equation}
\label{dgs}
\Delta_{\rm GS}^{\rm ED}=3.13816.
\end{equation}
These values of the gaps confirm the perturbative results of Sec.
\ref{sec:pert} saying that the lowest energy excitations are the ones
of the classical spins $r_{i}$, $s_{i}$ and the excitation within the
ground subspace of the higher order of magnitude.

\begin{figure}[t!]
\includegraphics[width=1\columnwidth]{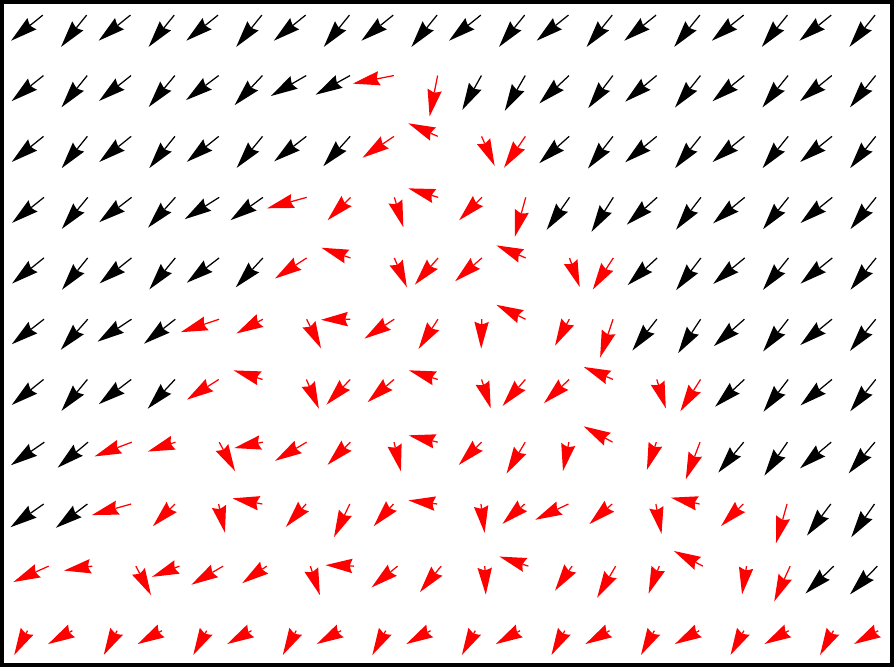}
\caption{(Color online)
Local spin averages $\left\langle \sigma_{2,i}^{x,z}\right\rangle $
and $\left\langle \sigma_{3,i}^{x,z}\right\rangle $ shown as arrows
in subspaces with highly entangled areas where both $r_{i}$'s and
$s_{i}$'s are negative (marked in red) for the system of the size
$L=10$ obtained via ED; spin disorder increases from top to bottom.
Every line corresponds with a different subspace, the first one with
the ground subspace and the last one with the highest excited one.
The horizontal (vertical) components of the vectors (arrows)
correspond to the $x$($z$) components of the spins $\sigma_{p,i}$.
\label{fig:enta_arrows}}
\end{figure}

Probably the most interesting feature of the POM that cannot be
captured within MF approaches is the spin configuration and entanglement
in the highly defected subspaces, i.e., the subspaces where in certain
range of $i$ both $r_{i}$ and $s_{i}$ are negative. In the extreme case
of all $r_{i}$ and $s_{i}$ being negative it is not even possible to
obtain conclusive MF results. In the ED approach we are free of such
problems so in Fig. \ref{fig:enta_arrows} we show the ground state
spin configuration for $L=10$ dimers in the subspaces with a growing
region of defects. The configurations are presented as the lines of
arrows such that the first line corresponds with a ground subspace
(no defects) and the last one with a fully defected subspace
($r_{i}=s_{i}=-1$ for all $i$). As we can see, the spins within the
defected region (the red ones) seem to be disordered and change very
rapidly from site to site. Some of them have even positive $x$ or $z$
components indicating the bonds that give positive contribution to the
total energy (see discussion in Sec. \ref{sec:mf}),
which implies strong frustration.

Interestingly, for defected region sizes $l$ that do not exceed $6$
dimers we observe a kind of regularity, a motif of four neighboring
spins that repeats in an approximate fashion when the number of dimers
in the defected region is even. This feature does not occur for larger
regions except for the fully defected subspaces where the translational
symmetry is present. In this case spins order regularly but the ordered
moments are much smaller and the difference between sublattices is more
pronounced than in the ground state configuration. Here the average
values for the $p=2$ spins are
$\langle\sigma_{2,i}^{x}\rangle\simeq-0.2985$ and
$\langle\sigma_{2,i}^{z}\rangle\simeq-0.5419$ that give the ordered
moment,
$m=\{\langle\sigma_{2,i}^x\rangle^2+\langle\sigma_{2,i}^z\rangle^2\}^{1/2}
=0.6187$. In the ground state these quantities are
$\langle\sigma_{2,i}^{x}\rangle_0\simeq-0.7256$,
$\langle\sigma_{2,i}^{z}\rangle_0\simeq-0.5846$, and $m_0=0.9318$.
In both cases the configurations exhibit a two-sublattice translational
invariant structure with the sublattices related by the interchange of
the $x$ and $z$ components of spins. Similarly to the 2D Kugel-Khomskii
model in the regime between the antiferromagnetic and ferromagnetic
phase \cite{Brz12}, the spins seem to prefer being perpendicular to
their neighbors but here because of the lower dimension this picture
is more distorted by quantum fluctuations. Quite remarkably this
classical view of perpendicular spins is realized by the quantum
observables, i.e., we observe that in the fully defected subspace all
the NN $\langle\sigma_i^z\sigma_j^z\rangle$ correlations and all the
NN $\langle\sigma_i^x\sigma_j^x\rangle$ ones are equal to zero.
This supports the picture of classical spin configuration where spins
on one sublattice point along the $z$ axis and on the other one ---
along the $x$ axis. However the smallness of the ordered moments and
the non-trivial angle between the spins in the configuration given
by the ED indicate that this state is more complex and potentially
highly entangled.

\begin{figure}[t!]
\includegraphics[width=1\columnwidth]{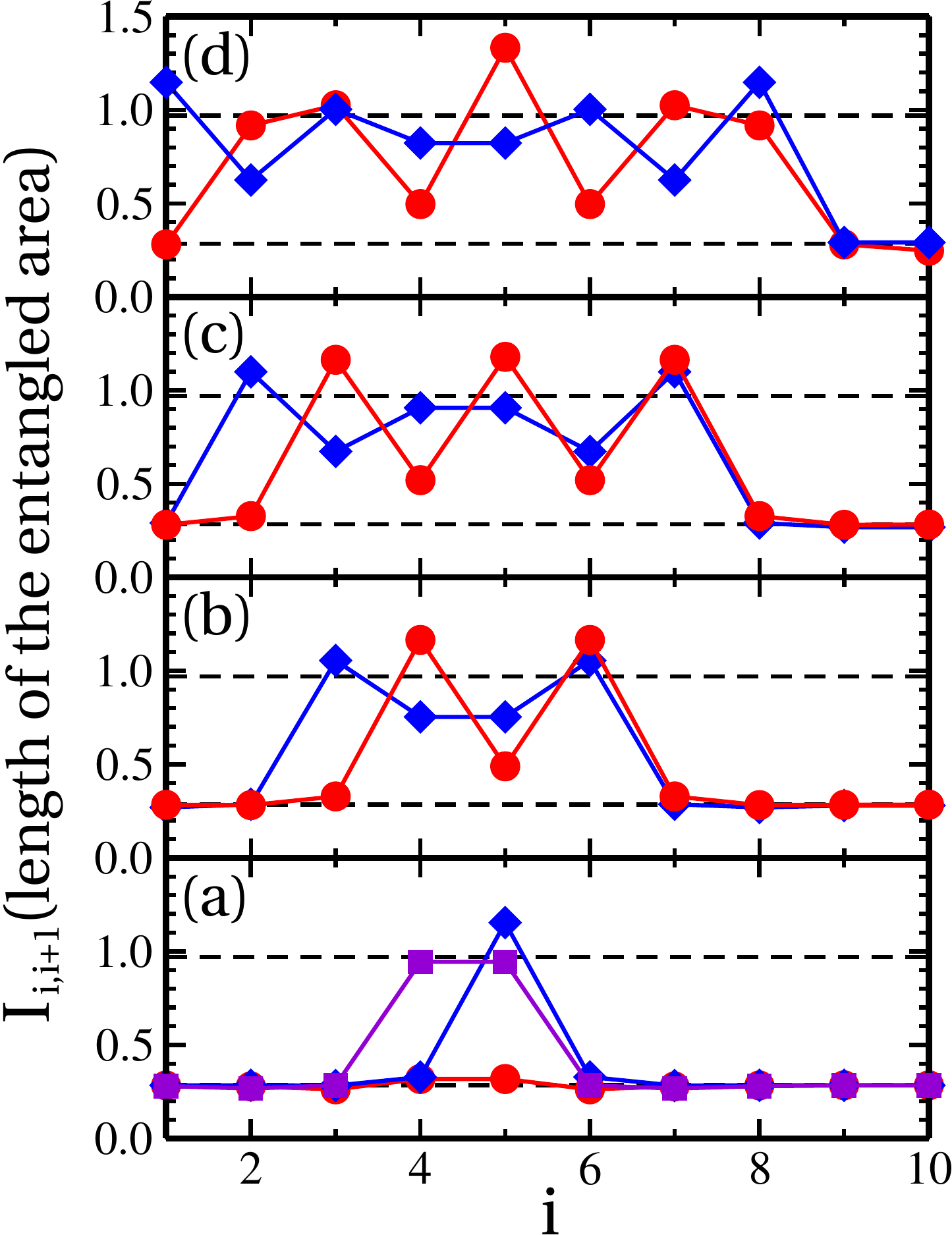}
\caption{(Color online) Mutual information of the NN sites $I_{i,i+1}$
for different sizes $l$ of the highly defected area (shown in Fig.
\ref{fig:enta_arrows}) as obtained for the chain of length $L=10$:
(a) $l=1$ (dots), $l=2$ (diamonds) and $l=3$ (squares),
(b) $l=4$ (dots), $l=5$ (diamonds),
(c) $l=6$ (dots), $l=7$ (diamonds), and
(d) $l=8$ (dots), $l=9$ (diamonds).
The dashed lines are the values of $I_{i,i+1}$ in the ground subspace
(bottom line) and in the totally defected subspace (upper line). }
\label{fig:mutual_10}
\end{figure}

To quantify the entanglement of the states described in Fig.
\ref{fig:enta_arrows} we will look at the mutual information
$I_{i,i+1}$ of the neighboring dimers in each of these states as
function of the site index $i$.
This quantity is defined by the von Neumann entropies of the dimers
$i$, $i+1$ and pair of dimers $\left\{ i,i+1\right\} $ as follows,
\begin{equation}
I_{i,i+1}={\cal S}_{i}+{\cal S}_{i+1}-{\cal S}_{i,i+1}
\end{equation}
where the von Neumann entropy ${\cal S}_{A}$ of any subsystem $A$ is
given by the formula,
\begin{equation}
{\cal S}_{A}=-\mbox{Tr}\rho_{A}\log_{2}\rho_{A},
\end{equation}
with $\rho_{A}$ being the reduced density matrix of the subsystem $A$
(i.e., we take the density matrix $\rho$ of the whole system and trace
it over all degrees of freedom outside the subsystem $A$).

In Fig. \ref{fig:mutual_10} we present the mutual information
$I_{i,i+1}$ for the states shown in Fig.
\ref{fig:enta_arrows} as function of $i$. The mutual information for
the lowest and highest subspaces does not depend on $i$ and is equal
to $I_{\rm GS}=0.28464$ and $I_{\rm HS}=0.96975$, respectively. These
values prove that the ground state in the fully defected subspace
is much more complex than the global ground state, as in the former
the entanglement between the neighboring dimers is roughly three times
stronger.

In the intermediate states that lie between the above two extremes the
mutual information in the defected areas is always bigger than outside
of them. This feature is very persistent in the sense that even if the
area free of defects contains only one dimer then the mutual information
of this dimer with respect to its neighbors is still roughly the same
as in the ground state, see Fig. \ref{fig:mutual_10}(d) --- this refers
of course also to other sizes of the defected area, compare Figs.
\ref{fig:mutual_10}(a), \ref{fig:mutual_10}(b), \ref{fig:mutual_10}(c)
and \ref{fig:mutual_10}(d). On the other hand, the mutual information
inside the defected areas behaves less regularly; we may say that it
has oscillatory character for the even sizes of the defected areas and
more plateau-like character for odd sizes. This however is only a
qualitative statement and probably larger systems should be studied to
determine some universal features of $I_{i,i+1}$ inside the defected
areas. What we can say for sure is that despite the observed
oscillations, $I_{i,i+1}$ never drops below the ground state level
$I_{\rm GS}$ in the defected areas although the value for the fully
defected subspace $I_{\rm HS}$ can locally exceed it.

\section{Summary and conclusions\label{sec:summa}}

We have shown a rather complete picture of the ground state and
low-energy excitations of the one-dimensional plaquette orbital model
defined by the Hamiltonian (\ref{eq:Ham}) using the perturbative,
mean-field and exact diagonalization approaches. First the model was
put in the block-diagonal form using spin transformation that reduces
the size of the Hilbert space by a factor of two. In this way we have
arrived at the model of interacting dimers consisting of the external
field terms acting on every site and the interaction terms having the
three-spin form, with signs given by the values of classical spins
resulting from the spin transformations or the eigenvalues of the local
symmetry operators. The perturbative approach has shown that the lowest
energy is obtained by setting all classical spins up and the lowest
excitations are obtained by creating defects in this polarized
configuration of classical spins.

The ground state configuration of the effective quantum spins is
characterized by the long-range order induced by the external field
acting along $-(x+z)$ direction. We have shown that the local average
values of these effective spins correspond with the average values of
the bonds in the initial ladder so the long-range spin-spin
correlations in the ground-subspace are the long-range bond-bond
correlations in the $Cx$-$Cz$ model. This resembles the N\'eel order
of the plaquettes energies found in the two-dimensional plaquette
orbital model \cite{Wen09}, however it has been shown that this is
an artifact of a deeper lying orientational order \cite{Bis10}.
The polarized ground state configuration of the effective spins is
slightly distorted by the quantum interaction terms that cause a
two-sublattice modulation of the order such that the sublattices are
related by the interchange of the $x$ and $z$ spin components.

In the lowest excited states the defects in the classical
spins cause an additional distortion in the configuration of quantum
spins through the local change of sign of the interaction terms. Such
change produces always local frustration of the interaction term that
in case of a single defect can be easily avoided by a local tilt of
the spins along either $x$ or $z$ axis. However, in case of the two
defects this is not always possible and the frustration can result in
the super-additive increase in the excitation energy.

The inhomogeneous mean-field approach shows that the frozen distortions
of the spin configuration found in the lowest excited states are very
local; due to the external field terms the system returns to its ground
state ordering at the distance of $3$ dimers. This is consistent with
the exact diagonalization results indicating that quantum fluctuations
have a short range character in the present model, as both the ground
state energy per dimer and the gap saturate extremely fast with the
increasing system size --- already for $L=12$ both quantities provide
excellent estimates for the values in the thermodynamic limit. This
follows from finite spatial range of three-spin entanglement in the
effective chain spin model
makes also the mean field approaches very successful, as shown in
Table \ref{tab:energies}. The energy gap remains finite for growing
system size with the best estimate being $\Delta=0.437271$ Eq.
\eqref{egs}, as obtained for $L=12$, and unlike in the 2D plaquette
model \cite{Wen09} the ground state is unique. The ground state energy
per dimer (or per one plaquette of the original model) is for this
system $E_{0}=-3.789718$, see Eq. \eqref{delta}.

The strong locality of the model can be attributed to the fact that
most of the bond operators of initial Hamiltonian are transformed
into the external field terms. For this reason we can conclude about
the behavior of the model from relatively small system sizes, in
certain analogy to the critical quantum chains with Potts interactions
\cite{Alc13}. This also makes the excitation within
a ground subspace very costly as in the zeroth order we need to flip
a spin against the external field to make an excitation. The
estimation for such energy obtained by exact diagonalization,
$\Delta_{\rm GS}=3.13816$, see Eq. \eqref{dgs}, shows that it does not
change much in the higher orders, at least not in the ground-subspace.

This not very exciting picture of mostly classical spin model found in
the ground state changes drastically when the defects in classical spin
configuration create frustration that cannot be avoided. This happens
when for a given dimer $i$ both variables $r_i$ and $s_i$ are negative
or both the $P_i^z$ and $P_i^x$ symmetries have negative eigenvalues.
As we have seen from the mean-field approach and the exact
diagonalization such a double defect produces a more severe distortion
in the configuration of quantum spins and costs more energy (as also
shown by the first order perturbation expansion) than these two defects
separated by more than one dimer. Thus we have studied the spin
configuration and the entanglement, characterized by the mutual
information $I_{i,i+1}$ of the neighboring dimers, for the subspaces
with such defects accumulated in the central part of the chain for a
growing number of defected dimers. We have found that within the
defected areas:
(i) spins form a very irregular pattern that resembles a spin-glass
state, and
(ii) the mutual information $I_{i,i+1}$ is strongly increased with
respect to its values outside the area.
We note that this phenomenon is analogous to increasing entanglement
entropy when disorder increases in quantum critical chains \cite{San06}.

There are two subspaces that are exceptional --- the ground subspace
where $I_{i,i+1}$ is (on average) minimal and equal to
$I_{\rm GS}=0.28464$, and the fully excited subspace with $r_i=s_i=-1$
where $I_{i,i+1}$ is (on average) maximal and equal to
$I_{\rm HS}=0.96975$. In both of these subspaces the ground states
exhibit a two-sublattice long-range order, however in the latter the
ordered moments are much smaller than in the former and the neighboring
spins tend to be perpendicular to each other, i.e., the bond spin
correlations vanish. The behavior of the mutual information $I_{i,i+1}$
in the intermediate subspaces is quite remarkable; no matter how large
the defected area is, the mutual information for the dimers outside this
area is always small and very close to $I_{\rm GS}$ - this also applies
to the case when only one dimer is outside. On the other hand, on
crossing the border of the defected area $I_{i,i+1}$ jumps immediately
above $I_{\rm HS}$ and behaves in an oscillatory way within the area,
remaining larger than $I_{\rm GS}$.

To conclude, we have constructed a simple pseudospin model where it
is possible to obtain large areas of disorder (or a spin-glass-like
behavior) and entanglement embedded in rather classically ordered
surrounding only by tuning the values of the symmetry operators.
We believe that this is of interest for constructing future
quantum computing devices and the model could be realized by the
superconducting lattices of Josephson junctions.

\acknowledgments

We warmly thank Karol \.Zyczkowski for insightful discussions.
We kindly acknowledge support by the Polish National Science
Center (NCN) under Project No. 2012/04/A/ST3/00331.

\appendix

\section{Backward spin transformation}
\label{sec:back}

Having explicit form of the spin transformations given by Eqs.
(\ref{eq:Xtilde}), (\ref{eq:Ztilde}), (\ref{eq:Xprim}), (\ref{eq:Zprim})
and (\ref{eq:sig}), it is straightforward to find a direct tranformation
from the new degrees of freedom,
$\left\{ \sigma_{i,2}^{z},\sigma_{i,3}^{z},r_{i},s_{i}\right\} $,
to the old ones, $\left\{ Z_{i,1},Z_{i,2},Z_{i,3},Z_{i,4}\right\} $,
i.e.,
\begin{eqnarray}
\sigma_{i,3}^{x} & = & X_{i,3}X_{i,4},\nonumber \\
\sigma_{i,2}^{x} & = & X_{i,2}X_{i,3},\nonumber \\
\sigma_{i,3}^{z} & = & Z_{i,1}Z_{i,4},\nonumber \\
\sigma_{i,2}^{z} & = & Z_{i-1,1}Z_{i,2}.
\label{eq:sigxz_XZ}
\end{eqnarray}
The list is completed by the already known relations,
\begin{eqnarray}
r_{i}&=&Z_{i,1}Z_{i,2}Z_{i,3}Z_{i,4}, \nonumber \\
s_{i}&=&X_{i+1,2}X_{i,1}X_{i,4}X_{i+1,3}.
\end{eqnarray}

\section{Duality of the interaction \\ and the free term}
\label{sec:dual}

It is easy to notice that the form of the spin interaction in Eq.
(\ref{eq:Ham_sig}) that the three-spin terms behave as new Pauli
operators, in the sense that they satisfy all the canonical
commutation relations. Thus we can define new spins, $\tau_{i,p}$
as follows,
\begin{eqnarray}
\tau_{i,2}^{x} & = &
s_{i-1}\left(\sigma_{i-1,2}^x\sigma_{i-1,3}^x\right)\sigma_{i,2}^x,
\nonumber \\
\tau_{i,3}^{z} & = &
r_{i+1}\sigma_{i,3}^z\left(\sigma_{i+1,2}^z\sigma_{i+1,3}^z\right).
\label{eq:tauxz}
\end{eqnarray}

Here we just took the interaction terms from Eq. (\ref{eq:Ham_sig}) or
at those shown in Fig. \ref{fig:ham_sig}. However, the algebra is not
complete yet, we need to define the $z$ and $x$ counterparts of the
$\tau_{i,2}^{x}$ and $\tau_{i,3}^{z}$ operators. One can easily check
that these definitions should be,
\begin{eqnarray}
\tau_{i,2}^{z} &=&
r_{i+1}\sigma_{i,2}^{z}\left(\sigma_{i+1,2}^{z}\sigma_{i+1,3}^{z}\right),
\label{eq:tauzx}\nonumber\\
\tau_{i,3}^{x} &=&
s_{i-1}\left(\sigma_{i-1,2}^{x}\sigma_{i-1,3}^{x}\right)\sigma_{i,3}^{x}.
\end{eqnarray}
Having them one can transform ${\cal H}$ to find
\begin{eqnarray}
{\cal H} & = & \sum_{i=1}^{L}\left\{\,\tau_{i,2}^{x}+\tau_{i,3}^{z}\right.
\nonumber \\
 & + & r_{i}\tau_{i-1,3}^{z}\left(\tau_{i,2}^{z}\tau_{i,3}^{z}\right)
 +s_{i}\left(\tau_{i,2}^{x}\tau_{i,3}^{x}\right)\tau_{i+1,2}^{x}\nonumber \\
 & + & \left.r_{i}\tau_{i-1,2}^{z}\left(\tau_{i,2}^{z}\tau_{i,3}^{z}\right)
 +s_{i}\left(\tau_{i,2}^{x}\tau_{i,3}^{x}\right)\tau_{i+1,3}^{x}\,\right\} .
 \label{eq:Ham_tau}
\end{eqnarray}

\begin{figure}[b!]
\includegraphics[width=0.80\columnwidth]{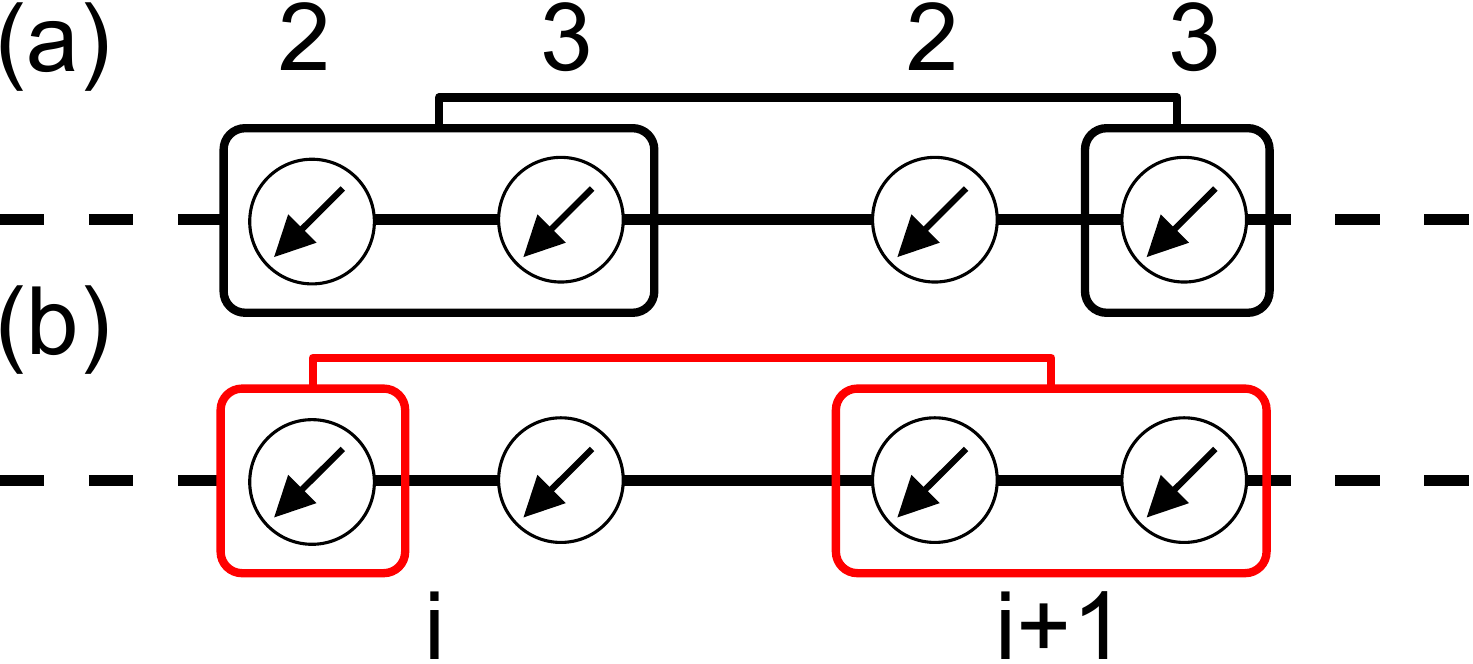}
\caption{(Color online)
Schematic view of the interactions in Eq. (\ref{eq:Hinv})
complementary to the ones present in Eq. (\ref{eq:Ham_sig}) ---
see Fig. \ref{fig:ham_sig}:
(a) the $x$ interactions (black frames), and
(b) the $z$ interactions (red frames). }
\label{fig:cmp_int}
\end{figure}

This Hamiltonian has a very similar structure to the one of Eq.
(\ref{eq:Ham_sig}), i.e., we have linear terms in $\tau_{i,p}$ and cubic
interaction terms with signs given by $r_{i}$ and $s_{i}$. There is
also a subtle difference as we get two more interaction terms [third
line of Eq. (\ref{eq:Ham_tau})] compared to the one already present in
Eq. (\ref{eq:Ham_sig}) but lose two of the linear terms. It is
straightforward to check that the structure of the two Hamiltonians is
exactly the same if we add to the Hamiltonian Eq. (\ref{eq:Ham_sig})
interaction terms of the complementary form,
$r_{i}\sigma_{i-1,2}^{z}\left(\sigma_{i,2}^{z}\sigma_{i,3}^{z}\right)$
and
$s_{i}\left(\sigma_{i,2}^{x}\sigma_{i,3}^{x}\right)\sigma_{i+1,3}^{x}$,
see Fig. \ref{fig:cmp_int}. In the other words, the Hamiltonian
${\cal H}_{\rm inv}$ of the form,
\begin{eqnarray}
{\cal H}_{\rm inv} & = & \sum_{i=1}^{L}\left\{
\left(\sigma_{i,2}^{z}+\sigma_{i,3}^{z}\right)+\left(\sigma_{i,2}^{x}
+\sigma_{i,3}^{x}\right)\right. \nonumber \\
& + & r_{i}\left(\sigma_{i-1,2}^{z}+\sigma_{i-1,3}^{z}\right)
\left(\sigma_{i,2}^{z}\sigma_{i,3}^{z}\right) \nonumber \\
& + & \left.s_{i}\left(\sigma_{i,2}^{x}\sigma_{i,3}^{x}\right)
\left(\sigma_{i+1,2}^{x}+\sigma_{i+1,3}^{x}\right)\right\},
\label{eq:Hinv}
\end{eqnarray}
is invariant under spin transformations (\ref{eq:tauxz}) and
(\ref{eq:tauzx}).

\end{document}